\documentclass[11pt]{article}
\pdfoutput=1
\usepackage{jheppub}

\usepackage{slashed}
\usepackage{amsmath,amssymb,graphicx} 
\usepackage{epsf,color}
\usepackage{hyperref}
\usepackage{cancel}
\usepackage{framed}
\usepackage{hyperref}

\definecolor{nicered}{rgb}{0.5,0.,0.}
\definecolor{nicegreen}{rgb}{0.,0.5,0.}
\definecolor{niceblue}{rgb}{0.,0.,0.5}
\definecolor{cerulean}{rgb}{0., 0.62,0.9}
\hypersetup{colorlinks,citecolor=nicegreen,linkcolor=nicered,urlcolor=niceblue}
\numberwithin{equation}{section}
\newcommand{\beq}{\begin{equation}}
\newcommand{\eeq}{\end{equation}}
\newcommand{\bea}{\begin{eqnarray}}
\newcommand{\eea}{\end{eqnarray}}
\newcommand{\bear}{\begin{eqnarray}}
\newcommand{\eear}{\end{eqnarray}}

\newcommand{\ba}{\begin{array}}
\newcommand{\ea}{\end{array}}

\begin{document}

\title{Dark Scalars and Heavy Neutral Leptons at DarkQuest }

\author[a]{Brian Batell,}

\author[b]{Jared A.~Evans,}

\author[c,d]{Stefania~Gori,}

\author[a]{Mudit Rai}

\affiliation[a]{Pittsburgh Particle Physics, Astrophysics,
and Cosmology Center, Department of Physics and Astronomy, University of Pittsburgh, Pittsburgh, USA}
\affiliation[b]{Department of Physics, University of Cincinnati, Cincinnati, Ohio 45221, USA}
\affiliation[c]{Santa Cruz Institute for Particle Physics, University of California, Santa Cruz, CA 95064, USA}
\affiliation[d]{Department of Physics, 1156 High St., University of California Santa Cruz, Santa Cruz, CA 95064, USA}

\emailAdd{batell@pitt.edu}
\emailAdd{jaredaevans@gmail.com}
\emailAdd{sgori@ucsc.edu}
\emailAdd{MUR4@pitt.edu}

\abstract{
The proposed DarkQuest beam dump experiment, a modest upgrade to the existing SeaQuest/SpinQuest experiment, has great potential for uncovering new physics within a dark sector. 
We explore both the near-term and long-term prospects for observing two distinct, highly-motivated hidden sector benchmark models:  heavy neutral leptons and Higgs-mixed scalars.   
We comprehensively examine the particle production and detector acceptance at DarkQuest, including an updated treatment of meson production, and light scalar production through both bremsstrahlung and gluon-gluon fusion.  In both benchmark models, DarkQuest will provide an opportunity to probe previously inaccessible interesting regions of parameter space on a fairly short timescale when compared to other proposed experiments. 
}

\maketitle

\allowdisplaybreaks

\section{Introduction} \label{sec:intro}

The hypothesis of a light, weakly coupled `dark'  or `hidden'  sector has received considerable attention in recent years.
 Though neutral under the Standard Model (SM) gauge group, dark sectors may exhibit rich dynamics, such as new forms of matter, new dark symmetries and forces, confinement, or spontaneous symmetry breaking, that could address some of the deficiencies of the SM. 
For example, the dark matter may be part of such a sector, communicating with the visible sector through a weakly coupled mediator, or the neutrino mass generation could be connected to new gauge singlet fermions within a dark sector. 

A vibrant experimental program to search for light weakly coupled particles has emerged over the last decade and promises to be a fertile area of research for many years to come; for a recent summary of existing and planned efforts, see the community studies \cite{Alexander:2016aln,Battaglieri:2017aum,Beacham:2019nyx,Alemany:2019vsk}.    
Among the critical components of this program, particularly in exploring GeV scale dark states, are proton beam fixed target experiments~\cite{Gorbunov:2007ak,Batell:2009di,Essig:2010gu}.
In these experiments, an intense proton beam impinges on a target, producing a torrent of SM particles alongside a smaller flux of relativistic dark sector particles. 
Due to their suppressed coupling to the SM, once produced these dark particles can travel macroscopic distances before decaying downstream into visible particles.
Given a suitable detector apparatus, the visible decay products can then be identified, characterized, and discriminated from potential background sources, 
which provides a promising means to probe and discover new light weakly coupled states. 

One particularly promising experiment is DarkQuest, a mild augmentation of the SeaQuest and SpinQuest experiments~\cite{Aidala:2017ofy}. The proposed DarkQuest upgrade entails the addition of an electromagnetic calorimeter (ECAL) to the existing SeaQuest muon spectrometer, which will extend the physics capabilities of the experiment.
These new capabilities will allow for DarkQuest to produce a suite of sensitive searches for dark particles decaying to a wide variety of SM final states such as electrons, muons, charged hadrons, and photons~\cite{Gardner:2015wea,Berlin:2018tvf,Berlin:2018pwi,Choi:2019pos,Dobrich:2019dxc,Tsai:2019mtm,Darme:2020ral}.
The experiment's high luminosity coupled with its short baseline would allow for sensitivity to both fairly short-lived particles ($c\tau\lesssim 1\, \rm m$) and more weakly-coupled particles with fairly low production rates.
Although a variety of other experimental proposals targeting dark sectors exist, DarkQuest is exceptional because most of the detector and infrastructure currently exists, is one of the few beam dump experiments with access to a high energy proton beam, would have an impressive range of sensitivity, and could provide novel results in comparatively short timescale.

In this work, we will study the potential sensitivity of DarkQuest to two highly motivated dark sector particles -- dark scalars and heavy neutral leptons (HNLs).  
Dark scalars that mix through the Higgs portal provide one of the simplest extensions the SM and may be connected to a variety of puzzles such as dark matter~\cite{Krnjaic:2015mbs}, inflation~\cite{Bezrukov:2009yw}, and naturalness~\cite{Graham:2015cka}. 
Heavy neutral leptons (also called right-handed neutrinos or sterile neutrinos) are strongly motivated by the observation of neutrino masses~\cite{Minkowski:1977sc,Yanagida:1979as,GellMann:1980vs,Glashow:1979nm,Mohapatra:1979ia,Schechter:1980gr} and GeV-scale HNLs may also play a role in the generation of the matter-antimatter asymmetry~\cite{Akhmedov:1998qx,Asaka:2005pn}. 
As we will demonstrate, DarkQuest has excellent prospects to explore substantial new regions of parameter space in these scenarios. Along with previous studies targeting a variety of dark sector models~\cite{Gardner:2015wea,Berlin:2018tvf,Berlin:2018pwi,Choi:2019pos,Dobrich:2019dxc,Tsai:2019mtm,Darme:2020ral}, our results lend further strong motivation for the DarkQuest ECAL upgrade, which will provide the basis for a rich and exciting experimental search program in the coming 5-10 years. 

The paper is organized as follows. In Section~\ref{sec:DarkQuestExperiment} we provide an overview of DarkQuest along with a general discussion of the methodology used in our sensitivity estimates. 
In Section~\ref{RHN} we consider the prospects for HNLs searches at DarkQuest, while searches for dark scalars are covered in Section~\ref{Scalar}. We present a summary of our results in Section~\ref{sec:summary}. In the Appendix, we provide the details about our calculation of dark scalar production.

\section{The DarkQuest Experiment}\label{sec:DarkQuestExperiment}

\begin{figure}[h]
\begin{center}
\includegraphics[width=.89\textwidth]{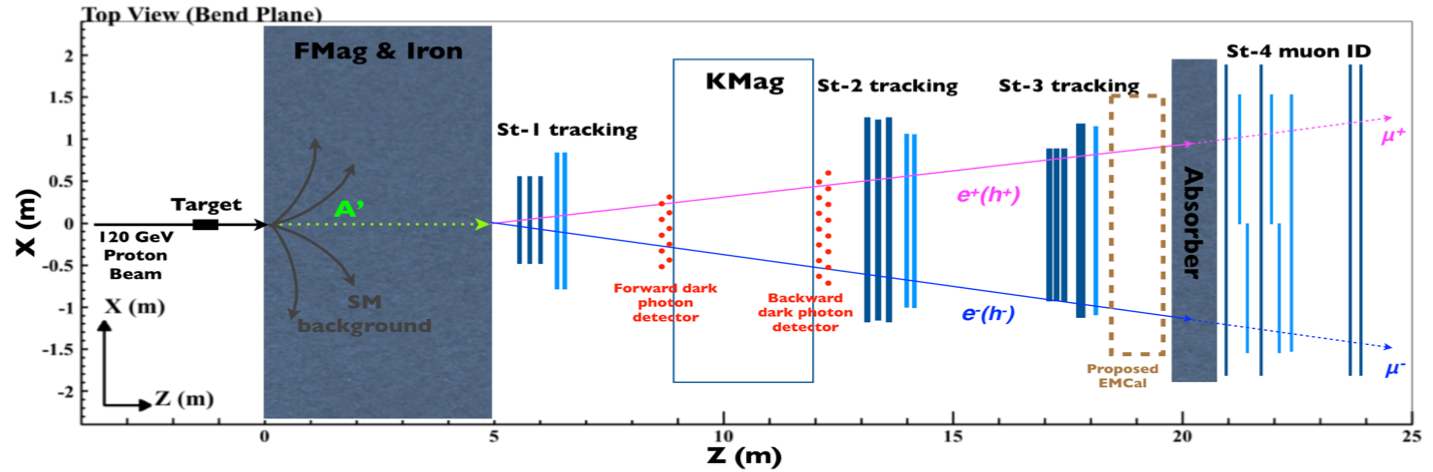}
\end{center}
\caption{Layout of the DarkQuest experiment. The SeaQuest experiment has the same layout, except for the ECAL (dashed brown region located near $z\sim19$ m) \cite{talkbySho}. \label{Fig:SeaQuest}}\end{figure}

The E906/E1039 SeaQuest/SpinQuest experiment is a proton fixed target beam dump spectrometer experiment on the neutrino-muon beam line of the Fermilab Accelerator Complex~\cite{Aidala:2017ofy}. A schematic layout of the experiment is shown in Figure~\ref{Fig:SeaQuest}. A high-intensity beam of 120 GeV protons (center of mass energy $\sqrt s\simeq15$ GeV) is delivered to a thin nuclear target. The target is situated $\sim 1$ m upstream of a 5 m long, closed-aperture, solid iron dipole focusing magnet (``FMAG"), which magnetically deflects soft SM radiation and also functions as a beam dump for the large majority of protons that do not interact in the target. This effectively allows only high energy muons, neutral kaons, and neutrinos to traverse the FMAG. The spectrometer consists of a high precision tracking system (St-1/2/3 tracking) and a muon identification
system (absorber and St-4 muon ID). 
An additional 3 m long open-aperture magnet (``KMAG'') is positioned at $z=(9-12)$ m and delivers a transverse momentum impulse of $\Delta p_T^{\rm{KMAG}}\sim 0.4$ GeV, enabling accurate momentum reconstruction of charged particles. 
In addition, in 2017 displaced vertex trigger hodoscopes were installed on both sides of the KMAG (see Figure~\ref{Fig:SeaQuest}), allowing for the detection of muons originating from the decays of exotic light long-lived particles after the dump. The experiment has been approved to collect $\sim 10^{18}$ protons on target in the coming two years, until 2023. 

On the horizon, there are plans to install a refurbished electromagnetic calorimeter (ECAL) from the PHENIX experiment~\cite{Aphecetche:2003zr} between St-3 and the absorber wall (see brown region in Figure~\ref{Fig:SeaQuest}). This will allow the upgraded experiment, DarkQuest, to search for a much broader set of dark sector displaced signatures, including electrons, charged pions and kaons, and photons. 
The DarkQuest experiment has a relatively compact geometry, making it well-suited to search for dark particles with $\mathcal O(10\,{\rm{cm}}-1\,{\rm{m}})$ lifetimes that are currently hidden to previous beam dump experiments with a much longer baseline.

Additional possible upgrades of the experiment (``LongQuest'') have been also proposed \cite{Tsai:2019mtm}. This includes additional trackers and calorimeters after station 4 of the SeaQuest spectrometer. 

The ultimate detectability of long lived dark particles at DarkQuest depends on several key factors. These include the production rate and kinematical properties of dark particles, their decay properties including branching ratios to final states containing charged particles and lifetime, the detector acceptance, and any potential SM background processes. In the remainder of this section we provide a brief discussion of these issues, which will motivate us to define two distinct run scenarios to be used later in our sensitivity projections for HNLs and dark scalars.   

\subsection{DarkQuest luminosity scenarios, Phase I and Phase II} \label{sec:particlesatDQ}

At DarkQuest both HNLs and dark scalars can be produced in meson decays (e.g., $K$, $D$, and $B$ mesons), while scalars can also be produced in the primary proton interactions through the proton bremsstrahlung and gluon  fusion processes. 
Assuming every proton interacts in the dump, an estimate of the effective integrated luminosity at DarkQuest is given by\footnote{An earlier study \cite{Berlin:2018pwi} used the effective luminosity for proton-proton collision within a single nuclear collision length of iron, $35{\rm{ab}}^{-1}\left(\frac{N_p}{1.44\times 10^{18}}\right)$.}
\begin{equation} 
\mathcal L \simeq \frac{ N_p }{\sigma_{p N}} \simeq \frac{ N_p A}{\sigma_{p {\rm Fe}}} \simeq N_p A\, \lambda_{{\rm int}} \,  \frac{\rho \, N_A}{A}  = 79  ~ {\rm ab}^{-1}  \left(\frac{ N_p}{10^{18}}\right),
\end{equation} 
where $N_p$ is the total number of protons on target, $\lambda_{\rm int} = 16.77$~cm~\cite{PDG-Mo} is the nuclear interaction length in iron, $\rho =  7.87$ g cm$^{-3}$ is the density of iron, and $N_A$ is the Avogadro's number.  In the second equality, we assume the per nucleon cross-section is the total cross-section on iron times the mass number, $A=56$.  A related quantity often seen in the literature is the the total hadronic cross per nucleon, which in iron is given by $\sigma_{p N}  \equiv  ( \lambda_{{\rm int}} \,  \rho \, N_A)^{-1}  \simeq 12.6$ mb. 

We will consider two benchmark luminosity scenarios in our projections below: a ``Phase I'' corresponding to $N_p = 10^{18}$ ($\mathcal L\sim79~$ab$^{-1}$ of integrated luminosity) which can be achieved on the couple of years time scale, and a ``Phase II'' scenario corresponding to  $N_p = 10^{20}$ ($\mathcal L\sim7.9~$zb$^{-1}$ of integrated luminosity)  which could potentially be collected over a longer time frame~\cite{Shiltsev:2017mle}. 

\subsection{Meson production at DarkQuest}

Given the considerable energy of the Main Injector protons and the substantial anticipated luminosity, mesons such as kaons, $D$-mesons, and $B$-mesons, as well as $\tau$-leptons, are abundantly produced at DarkQuest. 
Much of hidden sector particle production at DarkQuest thus occurs through the decays of these SM states.  Here we discuss our approach to modeling meson production at DarkQuest.

Kaons have an enormous production rate in primary proton collisions at DarkQuest, with an order one number of kaons produced per proton on target.  
However, since kaons are long lived and typically produced with boosts of order 10, their lab frame decay length is generally much longer than the characteristic hadronic interaction length, causing a significant attenuation of the kaon flux as they traverse the dump.   
Taking this into account, the number of kaons that decay before the first interaction length 
can serve as a useful proxy for the opportunities to produce hidden sector particles,  
\begin{equation}
N_{K_i {\rm  decay}}  \approx N_p \, n_{K_i} \, \Gamma_{K_i}\langle \gamma_K^{-1} \rangle \,  \lambda_{K} \,,  
\label{eq:NKs}
\end{equation} 
where $\lambda_{K} \approx 20$ cm is the kaon interaction length\footnote{Here we have assumed the kaon and pion interaction lengths in iron are similar and use the value given in Ref.~\cite{PDG-Mo}.},  
$n_{K_i} \sim 0.2$ is the number of kaons produced per proton on target at DarkQuest for each of $K^+$, $K^-$, $K^0_L$, and $K^0_S$, and $\langle \gamma_K^{-1} \rangle \sim 0.1$ is the mean inverse Lorentz boost. Both $n_{K_i}$ and $\langle \gamma_K^{-1} \rangle$  were estimated using {\textsc{PYTHIA 8}}~\cite{Sjostrand:2007gs}.  The values for $N_{K^\pm}$, $N_{K_L^0}$, and $N_{K_S^0}$ that decay before the first interaction length are shown in Table~\ref{table:SigmaMeson}. As expected, the number of $K_S^0$ is much larger than the number of $K_L^0$ and $K^\pm$ due to their much shorter lifetime. 

For $D$-meson production, we follow an approach that is similar to the one used by the SHiP experiment at CERN~\cite{SHiP:2018xqw}. 
We compute the $pp\to D^0,\bar D^0$ production cross section as a function of $\sqrt s$, using {\textsc{PYTHIA 8}}~\cite{Sjostrand:2007gs} with {\textsc{CTEQ6 LO}} parton distribution functions (PDFs) \cite{Pumplin:2002vw}.  We rescale these cross sections to match the cross sections measured in the interval $\sqrt s=(20-40)$ GeV \cite{CERN-SHiP-NOTE-2015-009,Lourenco:2006vw}. Using this rescaling, we estimate $\sigma (D^0,\bar D^0) \sim 1\,\mu$b at $\sqrt s=15$ GeV. Using the fragmentation
fractions for charm production, we obtain a charm production cross section $\sigma_{cc}=\sigma(D^0,\bar D^0)/f(c\to D^0)\sim 1.6\,\mu$b.  To estimate the fragmentation fractions, we generate hard $c\bar c$ processes in 
 {\textsc{PYTHIA 8}}~\cite{Sjostrand:2007gs} at the DarkQuest energy and extract the ratios. As a cross check, we have also used  {\textsc{PYTHIA 8}} to estimate the $B$ and $D$ fragmentation fractions at SHiP and LHC energies, finding relatively good agreement with the values quoted in Ref.~\cite{Bondarenko:2019yob}.  
The number of charm mesons produced for $N_p=10^{18}$ is shown in Table~\ref{table:SigmaMeson}  for $D^\pm$, $D^0$ and $\bar D^0$, and $D_s^\pm$.
 
\begin{table}[t]
\begin{center}
\begin{tabular}{|c|c||c|c || c|c || c|c|}
 \hline
  & $K$ mesons$^*$ &  & $D$ mesons & & $B$ mesons && Leptons\\
  \hline
  $K^\pm$ & $\sim1.8\times 10^{15}$ &   $D^\pm$ & $\sim6.8\times 10^{14}$ & $B^\pm$ & $\sim5.3\times 10^{7}$ &$\tau^\pm$ & $\sim4.7\times 10^{10}$\\
  $K_L^0$ & $\sim2.2\times 10^{14}$ &     $D_s^\pm$ & $\sim 2.0\times 10^{13}$ & $B_d,\bar B_d$ & $\sim5.3\times 10^{7}$ &$\tau^\pm_{D_s}$ & $\sim 1.1\times 10^{12}$ \\
  $K_S^0$ & $\sim1.2\times 10^{17}$ &       $D^0,\bar D^0$ & $\sim 1.3\times 10^{14}$ & $$ && &\\
  \hline
    \end{tabular}
      \caption{ Number of mesons and $\tau$s produced for $N_p = 10^{18}$.  For kaons, we present the number that decay before one nuclear interaction length, $\lambda_{\rm int}$, where the asterisk merely serves to flag that these are not the total amount produced. For taus, we present both those produced directly from electroweak interactions (first entry) and those from $D_s$ decays (second entry). The values shown are the sum of the production of the two mesons (e.g., particle + anti-particle).\label{table:SigmaMeson}}
        \end{center}
  \end{table}
 
We follow a similar procedure to compute the production rate of $B$-mesons. In Table~\ref{table:SigmaMeson}, we report the number of mesons produced for $N_p = 10^{18}$.  Due to $2 m_B + 2 m_p \sim \sqrt{s}$, there is substantial uncertainty on $\sigma_{bb}$ at DarkQuest beam energies.  In particular, Monte Carlo estimates with differing PDF choices can result in largely different values for the projected cross-section.  This can be primarily understood from the high uncertainty at large momentum fraction.  Unlike in the case of charm, we do not have empirical data to extrapolate from in a controlled manner.  Through exploring a variety of PDF choices, we found roughly an order of magnitude discrepancy for the projected cross-sections $\sigma(pp\to b\bar b) \sim 0.5 - 5$~pb.  Given this range, we choose $\sigma(pp\to b\bar b)=1$~pb throughout this work.  

In addition to meson decays, $\tau^\pm$ decays can produce dark sector particles.  At DarkQuest, the primary way of producing a $\tau$ lepton is through the decay of a $D_s$ meson with Br$(D_s\to \tau^\pm\nu_\tau) = (5.55\pm0.24)\%$~\cite{Tanabashi:2018oca}, which provides over an order of magnitude more $\tau$s than the direct electroweak production (see Table~\ref{table:SigmaMeson}, where the first entry represents the number of $\tau^\pm$ directly produced through electroweak processes).

We can compare the numbers in Table \ref{table:SigmaMeson} to the numbers obtained for higher energy proton beams as, for example, the 400 GeV SPS proton beam. The number of kaons \cite{Gorbunov:2020rjx}, $D$-mesons \cite{Bondarenko:2019yob}, and taus \cite{Buonaura:2016gho}  produced per proton on target is suppressed only by roughly an order of magnitude at the Fermilab Main Injector. A much larger suppression applies to $B$-meson production \cite{Bondarenko:2019yob}, for which the Main Injector loses roughly three orders of magnitude. 
For this reason, we generally expect DarkQuest to achieve a similar reach for dark sector states produced from light meson or tau decays.

Importantly, with the exception of $D$-mesons, most of these estimates consider only the particles produced in the incident protons primary interaction.  Secondary interactions of hard particles and beam remnants within the beam dump can also produce additional kaons and taus, which could potentially enhance the flux of dark particles.  The differential rates for these secondaries should be carefully evaluated in order for DarkQuest to most precisely state their sensitivity to a variety of models.  In this sense, our estimate of the reach should be considered conservative.

\subsection{Detector acceptance of DarkQuest}\label{sec:acceptance}

Next, we turn to the issue of the detector acceptance. Our considerations and approach to modeling the effect of the KMAG magnetic field and acceptance closely follows Ref.~\cite{Berlin:2018pwi}. 

A Monte Carlo simulation is used to compute the total detection efficiency. In particular, we will consider signal events to be those in which the dark particle decays to final states containing two quasi-stable charged particles  (i.e., electrons, muons, charged pions, and charged kaons) within a fiducial decay region at position $z \in (z_{\rm min}, z_{\rm max})$, located downstream of the FMAG. 
The daughter charged particles are then required to intersect tracking station 3, assumed to be a 2 m $\times$ 2 m square centered about the beam line and located approximately 18.5 m downstream of the dump (see Figure~\ref{Fig:SeaQuest}). We also model the effect of the KMAG magnetic field on charged particles trajectories by an instantaneous transverse momentum impulse of $\Delta p_T  = 0.4 \,{\rm  GeV} \times (\Delta z_K/3 {\rm m})$ applied in the $\hat x$ direction halfway through the particle's KMAG traverse, where $\Delta z_K$ is the distance traveled by the daughter particles through the KMAG\footnote{Note that Ref.~\cite{Berlin:2018pwi} applied the $p_T$ kick at the end of the KMAG.}. The total detection efficiency is then estimated according to~\cite{Berlin:2018pwi}
\beq 
\label{eq:efficiency}
\text{eff}=m\,\Gamma\,\int_{z_{\text{min}}}^{z_{\text{max}}}dz\,\sum_{\text{events \ensuremath{\in} geom.}}\frac{e^{-z\,(m/p_{z})\,\Gamma}}{N_{\text{\rm MC}}\,p_{z}}~,
\eeq
where $m$, $\Gamma$, and $p_z$ are the mass, width, and $\hat z$-component of the momentum of the dark particle, respectively. The sum in (\ref{eq:efficiency}) is carried out over those events falling within the geometric acceptance as described above, 
and $N_{\rm MC}$ represents the total number of simulated events. 

We will define two fiducial decay regions for our study that will be associated with our near future and long term run scenarios. As we will discuss in Secs. \ref{Sec:HNLAcceptance}, \ref{Sec:ScalarAcceptance}, the detection efficiency for the two fiducial decay regions is relatively sizable, ranging from $\sim$ few $\times 10^{-2}$ to $\sim 1$, depending on the particular production and decay mode of the dark particle.

For our Phase I scenario, we require that the dark particle decays within the 5~m~$-$~6~m region immediately downstream of the FMAG. 
The main advantages of this choice are that the charged daughter particles are tracked in Station I and their trajectories are bent by the KMAG magnetic field, making accurate momentum reconstruction feasible and greatly helping  with particle identification, vertex reconstruction, and background rejection.
 
For our Phase II scenario we will consider the longer fiducial decay region of 7~m~$-$~12~m. Given the higher luminosity in our Phase II scenario, we expect more background events, e.g., from $K_L^0$ particles which pass through the FMAG and decay semileptonically. 
As discussed in Ref.~\cite{Berlin:2018pwi}, these backgrounds could be further mitigated with additional shielding in the 5~m~$-$~7~m region, partially explaining the motivation of the 7~m~$-$~12~m fiducial region.  In addition, the 7~m~$-$~12~m fiducial region would increase 
the geometric acceptance.  While this choice allows for an appreciable enhancement of the overall signal rate and for additional suppression of backgrounds, it is not without additional challenges. 
For example, momentum reconstruction will be more challenging since the daughter particles would not pass through the first tracking station. 

Our benchmark scenarios discussed here should be considered as preliminary, and a dedicated study of the potential backgrounds and signal region optimization is warranted. The DarkQuest collaboration is 
currently investigating the several sources of backgrounds, 
with a focus on the $e^+e^-$ signature characteristic of dark photons 
\cite{BackGroundPrivate}. 
While awaiting a definitive study from the collaboration, a crude estimate suggests that it will be possible to observe signals over the $K_L^0$ decay  backgrounds. 
For the signatures investigated in this paper, the dominant sources come from the production of $K_L^0$ with subsequent semi-leptonic $K_L^0\to\pi^\pm e^\mp\nu$, $K_L^0\to\pi^\pm \mu^\mp\nu$ or purely hadronic $K_L^0\to\pi^+\pi^-\pi^0$,  $K_L^0\to\pi^+\pi^-$ decays. Roughly $10^{17}$ $K_L^0$ will be produced in the beam dump during Phase I. Taking the kaon interaction length in iron to be $\sim 20$ cm, we expect approximately
$\sim 10^6$ kaons to escape FMAG, and $\mathcal O(10^4)$ of which will decay in 5 m - 6 m. Accounting for branching ratios and geometric acceptance, we find that, depending on the particular final state, 
$\mathcal O(100 - 1000)$ $K_L^0$ will decay in the fiducial region with decay products detected by DarkQuest.
Despite the substantial increase in luminosity, the situation during Phase II can be much improved over Phase I provided additional shielding is in place between 5 m - 7 m. 
While approximately $\sim 10^{19}$ $K_L^0$ will be produced in Phase II, a similar estimate as given for Phase I suggests that depending on the specific final state, $\mathcal O(1-10)$ $K_L^0$ will traverse 7 m of iron, decay in the 7 m - 12 m fiducial region, and will lead to detectable decay products. 
Depending on the final state signature, additional handles can be utilized to further mitigate these backgrounds.
In Secs. \ref{RHNReach}, \ref{Sec:reachScalars}, we will estimate how many of these $K_L^0$ will result in background events for the several signatures. When discussing the DarkQuest reach for dark scalars and HNLs, we will require 10 signal events, but the true requirement against background may be more or less depending on the expected background population specific to the mass and decay paths.

\section{Heavy Neutral Leptons}\label{RHN}

Heavy neutral leptons (HNL), $\hat N_i$, can interact with the SM neutrinos through the neutrino portal operator 
\beq
-\mathcal L\supset \lambda_N^{ij} \, \hat  L_i  \, H  \hat N_j+{\rm{H.c.}}\,,
\eeq
where 
$H$ is the SM Higgs doublet and $\hat L_i=(\nu_i,\ell_i)^T$ is the SM lepton doublet of flavor $i$. 
Because of these operators, after electroweak symmetry breaking, the HNLs will mix with the SM neutrinos. 
We will refer to the unhatted fields $\nu_i$ and $N_i$ as the corresponding mass eigenstates of the light SM neutrinos and HNLs, respectively, and  the relation between the flavor and mass bases is described by a mixing matrix, $U$. The phenomenology of HNLs largely follows from their induced couplings to electroweak bosons, which in the limit of small mixing angles are given by
\begin{equation}
\label{eq:HNL-weak-interaction}
{\cal L} \supset \frac{g}{\sqrt{2}} U_{ij}  W_\mu^- \, \ell_i^\dag  \, \overline \sigma^\mu N_j  + \frac{g}{2 c_W} U_{ij} \,   Z_\mu \,   \nu_i^\dag \,\overline \sigma^\mu N_j  +{\rm H.c.}
\end{equation}
 Additionally, we will assume that $N$ is a Majorana particle throughout this work. Majorana HNLs are particularly motivated as they arise in the Type-I seesaw mechanism for neutrino mass generation. While the Type-I seesaw naively leads to mixing angles of parametric size $\sim \sqrt{m_\nu/m_N}$, which is extremely small for GeV-scale HNLs, we note that there are schemes such as the inverse seesaw~\cite{Mohapatra:1986aw,Mohapatra:1986bd,Bernabeu:1987gr} and linear seesaw~\cite{Malinsky:2005bi} where the mixing angles can be much larger. For the purposes of characterizing the DarkQuest sensitivity, we will take a phenomenological approach, as is commonly done in the literature, assuming the existence of a single HNL state, $N$, in the mass range of interest, which dominantly mixes with a particular neutrino flavor, i.e., dominant electron-, muon-, or tau- flavor mixing. 
In this case, the phenomenology is dictated by the HNL mass, $m_N$, and mixing angle, denoted by $U_e$, $U_\mu$, or $U_\tau$, respectively,  for the three mixing scenarios. 
 If these assumptions were relaxed, we expect the phenomenological implications relevant for DarkQuest are typically only slightly different than in a flavor-aligned case.

\subsection{HNL production}\label{RHNProduction}

\begin{figure}
\includegraphics[width=0.48\textwidth]{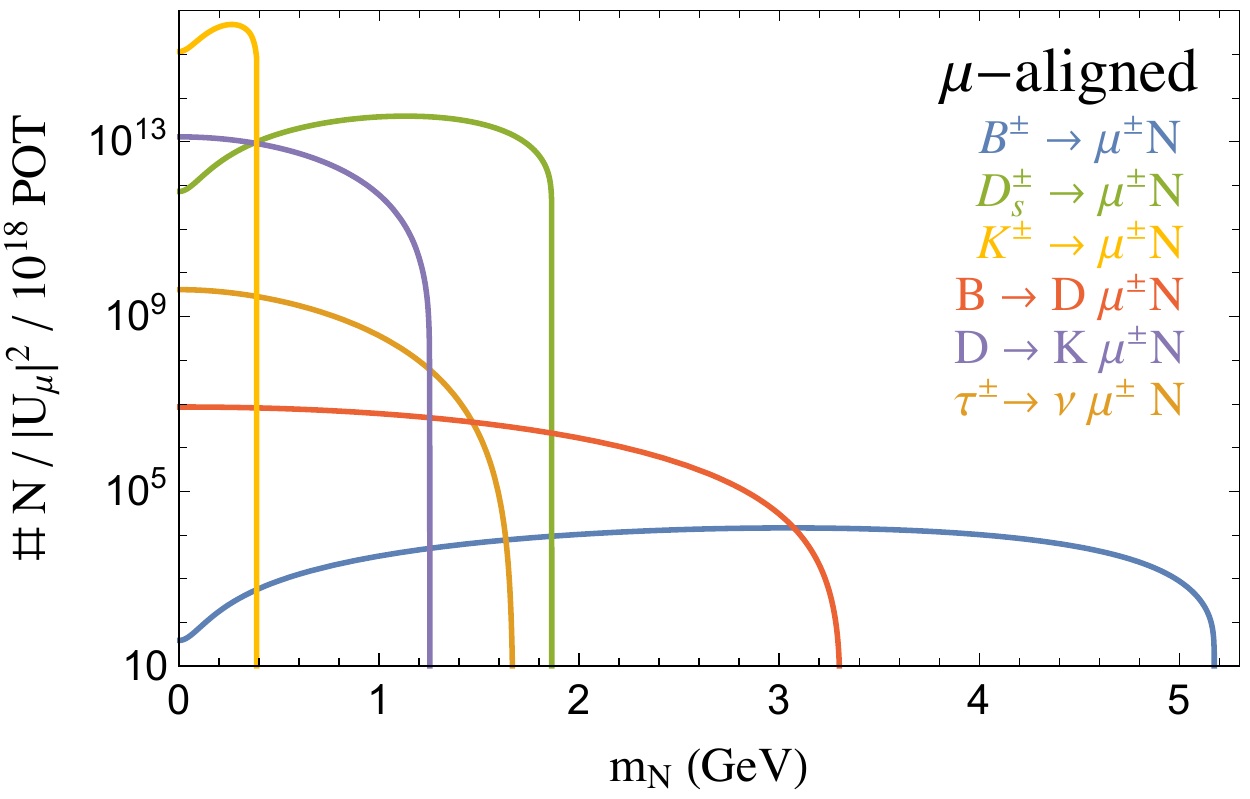}\hspace{3mm}
\includegraphics[width=0.48\textwidth]{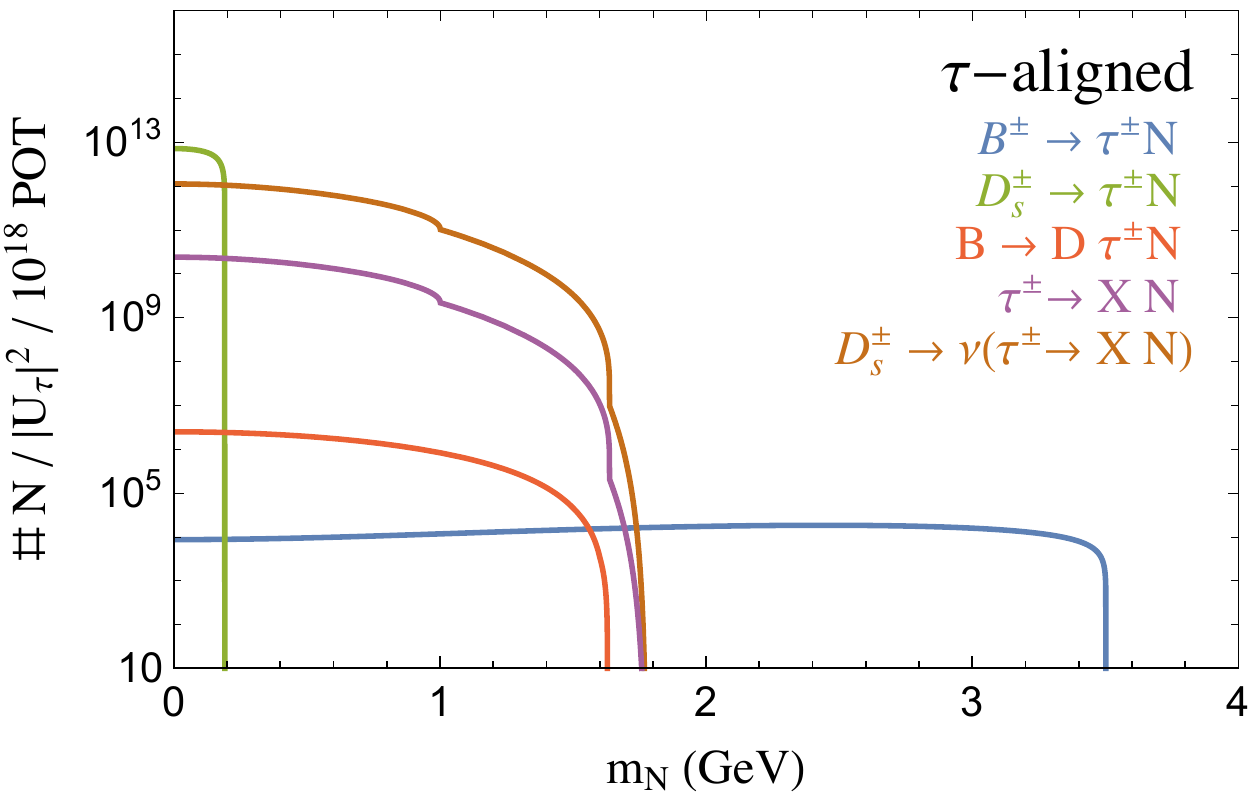}
\caption{Number of $\mu$-aligned (left) and $\tau$-aligned (right) HNLs produced through meson and lepton decays, using $10^{18}$ protons on target and mixing angle equal to 1. The $e$-aligned scenario is nearly identical to $\mu$-aligned one. For this reason, we do not show it here. The most important channels are $B^\pm\to \ell^\pm N$ (blue),  $D_s^\pm\to \ell^\pm N$ (green), $B$ mesons decaying to a charm meson and $\ell N$ (red, denoted as $B\to D \ell^\pm N$), $D$ mesons decaying to a strange meson and $\mu N$ (purple, denoted as $D\to K \mu^\pm N$, left figure only), $\tau^\pm \to \nu\mu^\pm N$ (yellow, left figure only), $\tau^\pm \to XN$ (purple, right figure only), and $D_s^\pm\to \nu (\tau^\pm \to XN)$ (brown, right figure only). \label{Fig:HNLProd}
}  
\end{figure}

As a consequence of the interactions in (\ref{eq:HNL-weak-interaction}), HNLs can be copiously produced at DarkQuest through the decays of mesons and $\tau$ leptons.  Meson and $\tau$ production at DarkQuest is discussed in Sec.~\ref{sec:particlesatDQ} and summarized in Table~\ref{table:SigmaMeson}. 
For example, HNLs can be produced in the two body decays of charged pseudoscalar mesons, $P \rightarrow \ell_i N$. In the regime $m_\ell \ll m_N \ll m_P$, the branching ratio is given by~\cite{Bondarenko:2018ptm}
\begin{equation}
{\rm Br}(P \rightarrow \ell_i N) \simeq  \tau_P \frac{G_F^2}{8\pi}\,  f_P^2 \,  m_P \, m_N^2 \, |V_{\alpha\beta}  |^2 \, |U_i|^2, 
\label{eq:P-decay}
\end{equation}
where $\tau_P$, $f_P$, and $m_P$ are the meson lifetime, decay constant, and mass, respectively, and the CKM matrix element, $V_{\alpha\beta}$, is dictated by the valence quark content of $P$ (e.g., $V_{cd}$ for $D^\pm$, etc.). The two body decay rates (\ref{eq:P-decay}) scale as $m_N^2$ as a consequence of the chirality flip, and are thus enhanced for heavier HNLs.

Three body decays of mesons to HNLs are also important and can even be the dominant production mechanism depending on the HNL mass. 
Although phase space suppressed, the three body meson decay rates do not suffer from the CKM or chirality flip suppressions characteristic of the two body decays in (\ref{eq:P-decay}).
HNLs can furthermore be produced through $\tau$ decays (e.g., two body decays involving hadronic resonances, or three body leptonic decays) and are subject to similar considerations.

For all meson and $\tau$ branching ratios, we use the expressions in Ref.~\cite{Bondarenko:2018ptm}.  
The total number of HNLs produced at DarkQuest through different pathways is summarized in Figure~\ref{Fig:HNLProd}, where we utilized a luminosity of $10^{18}$ protons on target.

\subsection{HNL decays}\label{RHNDecays}

\begin{figure}
\includegraphics[width=1\textwidth]{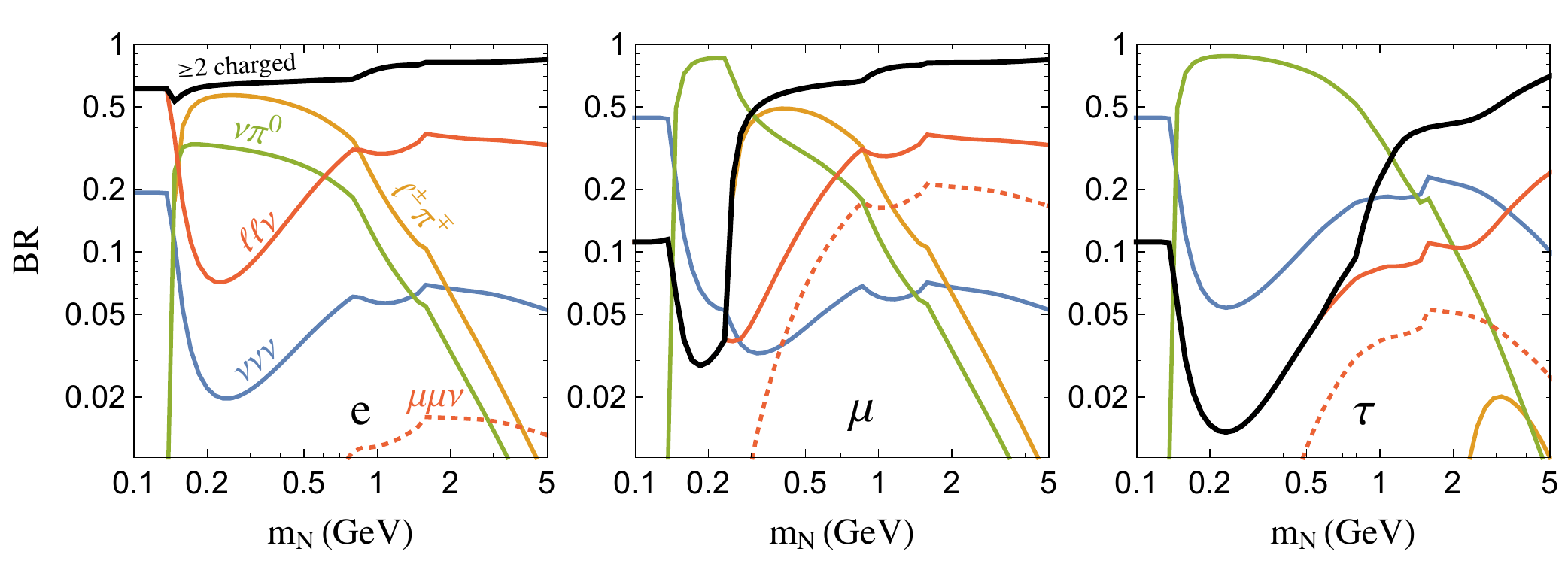}
\caption{Branching ratios of the HNLs. The three panels represent HNLs mixed either with the electron (left panel), muon (middle panel), or tau (right panel) neutrinos. In each figure, we show the branching ratios into three SM neutrinos $\nu\nu\nu$ (blue), $e^\pm\pi^\mp$ or $\mu^\pm\pi^\mp$ (gold), $\nu\pi^0$ (green), one neutrino and two charged leptons of any flavors (red), and one neutrino and two muons (dotted red). The thick black curve represents the sum of the branching ratios into two or more charged tracks. \label{Fig:BRRHN}
} 
\end{figure}

Once produced at DarkQuest, HNLs will decay through the weak interactions (\ref{eq:HNL-weak-interaction}) to a variety of SM final states. Since their decays proceed through an off-shell heavy electroweak boson, 
GeV-scale HNLs are generically long lived and can easily traverse the beam dump at DarkQuest before decaying. There is a rich variety of HNL decay modes,  including a pseudo-scalar meson and a lepton, a vector meson and a lepton, a lepton and two or more pions, or three leptons (including three neutrinos). We note that there is some disagreement in the literature about the corresponding rates. We have verified the results of Refs.~\cite{Bondarenko:2018ptm,Ballett:2019bgd}, and utilize these expressions for the neutrino decays. 

In Figure~\ref{Fig:BRRHN} we show the branching ratios of HNLs in the $e-$aligned, $\mu-$aligned, and $\tau-$aligned case (left, center, and right panel, respectively).  For HNL masses below 1.5 GeV, we determine the total hadronic rate as the sum of exclusive meson decay rates, while above 1.5 GeV, we switch to using the inclusive $N\to q\bar q' \ell$ rate, assuming exclusive rates are contained within this value.  As we can observe from the figure, the branching ratio into the invisible $\nu\nu\nu$ final state (in blue in the figure) is quite subdominant as long as the HNL has a mass above the pion mass. The other channels presented in the figure contain visible particles that are in principle observable by DarkQuest. The red dotted curve represents the decay into one neutrino and two muons. 
The corresponding branching ratio is also relatively suppressed, especially in the $e-$aligned, and $\tau-$aligned scenarios. This is the only channel that can be easily identified now by the SeaQuest experiment, without the ECAL upgrade. 

Provided the ECAL upgrade is installed, DarkQuest will have the capability to also search for a variety of HNL decays containing multiple charged particles  in addition to muons. 
Among all visible channels, the $\pi^0\nu$ channel is likely to be the most difficult one because of the challenging $\pi^0$ identification and large sources of backgrounds arising e.g., from the SM $K_L^0\to 3\pi^0, K_S^0\to\pi^0\pi^0$ processes, where some of the pions are missed or misidentified by the detector. For this reason, in the calculation of the DarkQuest reach on HNLs, we conservatively 
do not include this channel. The bold black line in Figure~\ref{Fig:BRRHN} shows the observable branching ratio used in this work, which is obtained by summing all branching ratios resulting in at least two charged particles. 

In estimating the sensitivity below we will require 10 signal events, working under the assumption that backgrounds can be brought down to the level of a few events. 
The FMAG, i.e., the  5 m magnetized beam dump, serves to mitigate most of the backgrounds by sweeping away charged particles and largely blocking the most dangerous neutrals.  
Several potential sources remain and the ultimate size of these is the subject of current study~\cite{BackGroundPrivate}. 
One of the most relevant backgrounds comes from $K_L^0$ particles that penetrate the dump and decay in the fiducial region. 
As we discussed in Sec. \ref{sec:acceptance}, we expect $\mathcal O(100-1000)$ of such $K_L^0$ in Phase I and $\mathcal O(1-10)$ in Phase II. 
The decay $K_L^0\to \pi^\pm e^\mp \nu$ will be background to the $N\to e^+e^-\nu$ and $N\to e^\pm\pi^\mp$ signatures presented in Fig. \ref{Fig:BRRHN}. 
For the former, a pion rejection factor of order $\sim 1\%$ will be sufficient to suppress the $K_L^0\to \pi^\pm e^\mp \nu$ background to $\mathcal O(10)$ ($< 1$) events for Phase I (Phase II). 
This level of electron-pion discrimination should be feasible with the planned ECAL upgrade~\cite{Aphecetche:2003zr}. 
 For the latter signal, the background could be suppressed  through suitable kinematic cuts such as a cut on the $m_{e\pi}$ invariant mass. However, a detailed study of these possibilities requires a careful modeling of $K_L^0$ production in the FMAG, which is beyond our current scope. 
 For signatures involving muons, the existing SeaQuest spectrometer already has the capability to distinguish muons, which pass through the absorber and are detected in the Muon-ID system (see Figure~\ref{Fig:SeaQuest}), from charged hadrons, which do not penetrate the absorber.  As above, muonic backgrounds to the $N\to \mu^\pm\pi^\mp$ signature can arise from decays such  $K_L^0\to \pi^\pm \mu^\mp \nu$, while the $N\to \mu^+\mu^-\nu$ channel should have very small backgrounds.

\subsection{Detector acceptance}\label{Sec:HNLAcceptance}

\begin{figure}
\begin{center}
\includegraphics[width=.5\textwidth]{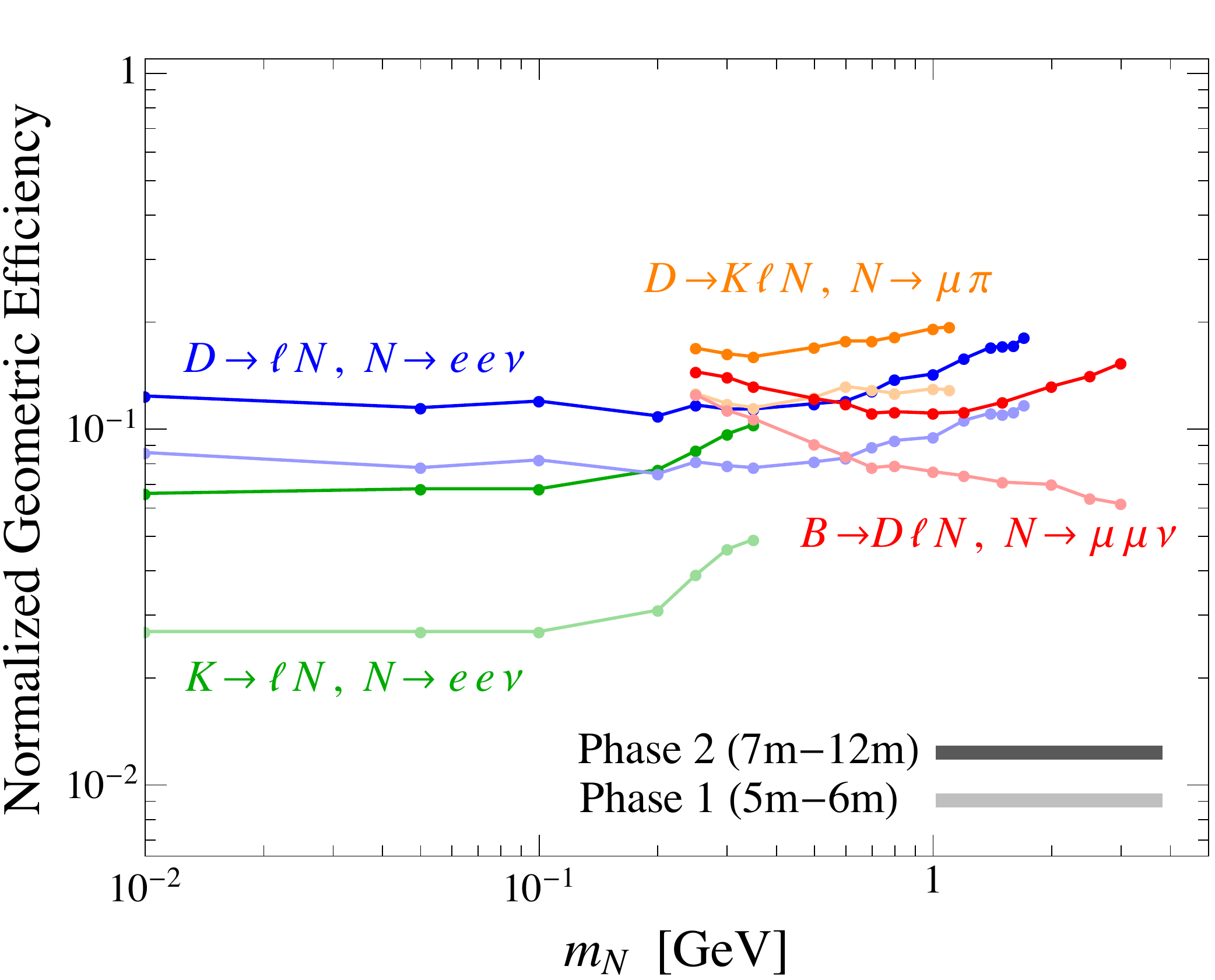}~~~~
\end{center}
\caption{
Geometric acceptance as a function of the HNL mass normalized to the number of HNLs decaying within the fiducial decay region in the large lifetime limit (i.e., the HNL decay length is much larger than the detector size).
We show separately the efficiency for HNLs that are produced and decay through several representative channels, including  $K\rightarrow \ell N, N \rightarrow e e \nu$ (green), $D\rightarrow \ell N$, $N\rightarrow ee \nu$ (blue), $D\rightarrow K \ell N, N\rightarrow \mu \pi$ (orange) and $B\rightarrow D \ell N, N\rightarrow \mu \mu \nu$ (red),
 and for two run scenarios: Phase II, 5m - 6m (lighter darker), Phase II, 7m - 12m, (darker color).  \label{Fig:HNL-eff}}
\end{figure}

We follow the procedure outlined in Sec.~\ref{sec:acceptance} to compute the geometric acceptance for HNLs at DarkQuest. 
To reduce the complexity for a clear presentation, we show in Figure~\ref{Fig:HNL-eff} the normalized geometric efficiency in the large lifetime limit.  
To compute these curves, we consider the $\mu$-aligned scenario and the large lifetime regime, i.e., we assume that the HNL decay length is much larger than the detector size so that the differential probability to decay is a constant with distance, and normalize to only the particles that decay within the fiducial region. This limit is relevant for small mixing angles. The different colored curves in Figure~\ref{Fig:HNL-eff} correspond to several representative production and decay modes of the HNL. The lighter (darker) curves represent the acceptance for Phase I (5 m -  6 m) (Phase II (7 m - 12 m)). 
Overall, the acceptance is relatively large, ranging from a few $\%$ to $\sim 20\%$ depending on the HNL production/decay mode, and is fairly constant with the HNL mass.
As expected, the acceptance for Phase I is somewhat smaller than that for Phase II, since for Phase II the HNLs decay typically closer to tracking station 3. 

\subsection{The DarkQuest reach for HNLs}\label{RHNReach}

With our estimates for HNL production, decays, and experimental acceptance in hand, we can compute the total number of signal events in the SM final state $i$ expected at DarkQuest according to  
\begin{equation}
N_{\rm signal} = N_{N} \times {\rm Br}_i \times {\rm eff}_i \, .
\label{eq:NHNL}
\end{equation}
Here $N_N$ is the number of HNLs produced in a given production channel (see Section~\ref{RHNProduction} and Figure~\ref{Fig:HNLProd}), 
${\rm Br}_i$ is the branching ratio for $N\rightarrow i$ (see Section~\ref{RHNDecays}, and Figure~\ref{Fig:BRRHN}), and ${\rm eff}_i$ is the experimental efficiency to detect the final state $i$, 
computed using (\ref{eq:efficiency}). 

\begin{figure}
\includegraphics[width=0.48\textwidth]{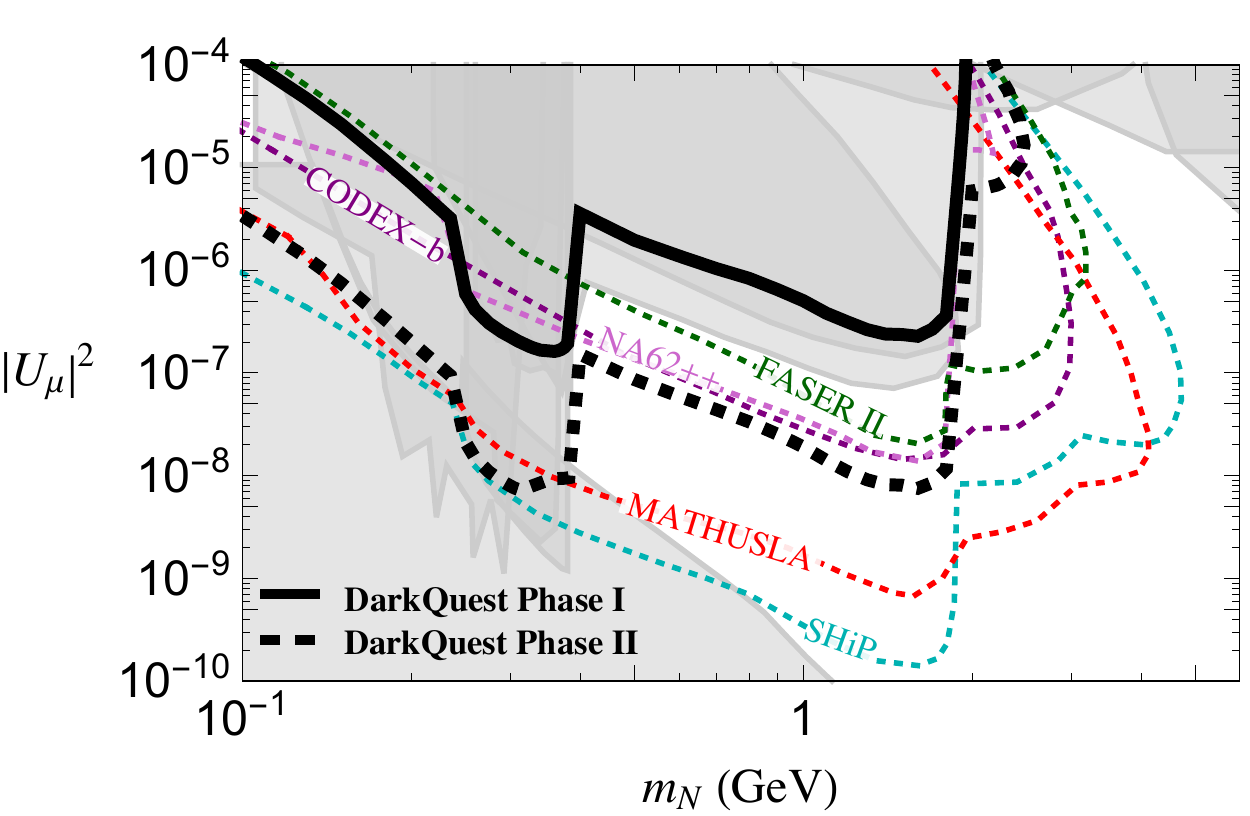}\hspace{3mm}
\includegraphics[width=0.48\textwidth]{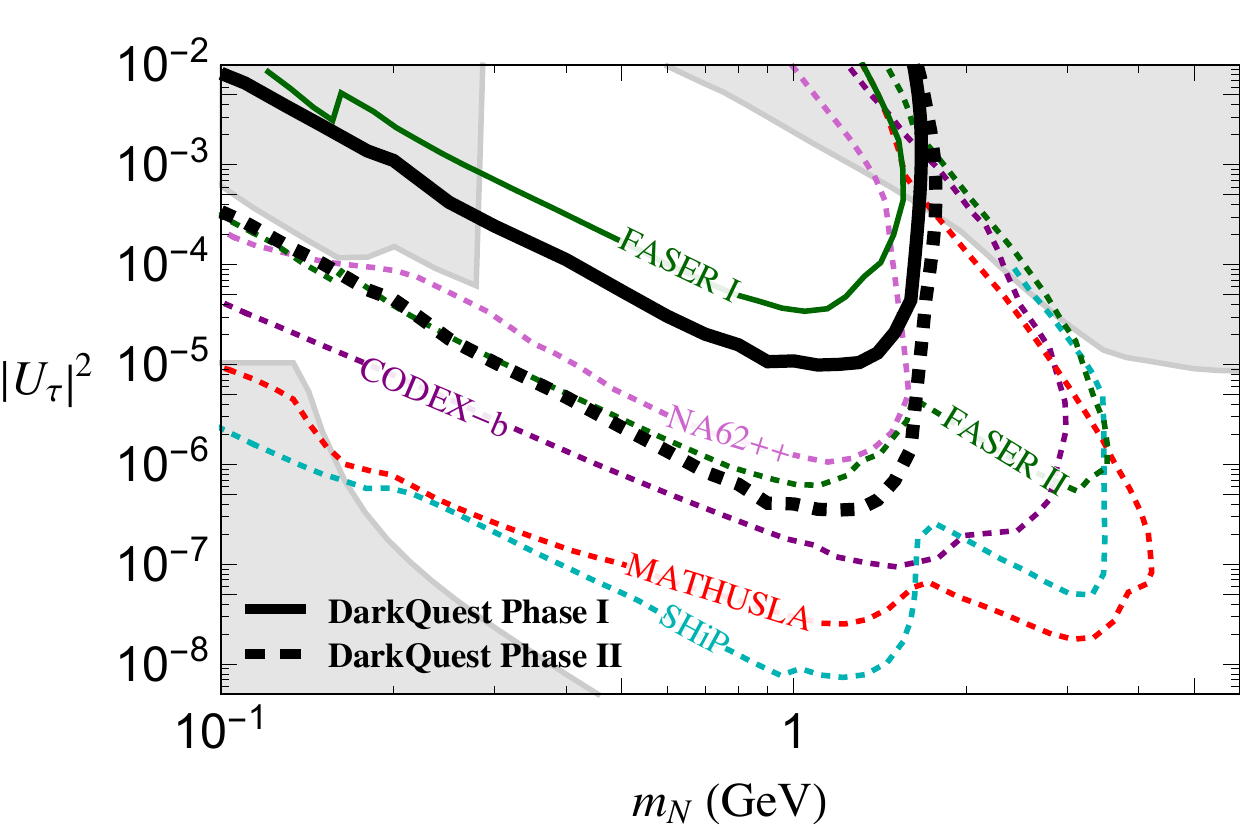}
\caption{Projected reach for $\mu$-flavored HNLs (left panel) and $\tau$-flavored HNLs (right panel) in the $m_N$ vs $\left|{U_{\mu,\tau}}\right|^2$ plane.  
DarkQuest Phase I is represented by the black solid line, and Phase II by the black dashed line. Current limits (gray) and limits from proposed future experiments (colored dashed) are also displayed for comparison; see the text for a details. Limits are set requiring 10 signal events.
\label{Fig:HNLReach}}
\end{figure}

A summary of the projected reach is shown in Figure~\ref{Fig:HNLReach} for $\mu$- and $\tau$-flavored HNLs decaying inclusively to final states containing two or more detected charged tracks. 
The solid black (dashed black) contour specifies the HNL mass - squared mixing angle parameters leading to 10 signal events according to (\ref{eq:NHNL}) for the Phase I (Phase II) run scenario. 
 We note that the projected reach for $e$-aligned HNLs is very similar to the $\mu$-aligned reach shown in Figure~\ref{Fig:HNLReach}.  For this reason, we do not show the $e$-aligned scenario in the figure.
 We also show in the shaded gray regions the existing experimental or observational limits, including  CHARM~\cite{Bergsma:1985is}, PS191~\cite{Bernardi:1987ek}, DELPHI~\cite{Abreu:1996pa}, NuTeV~\cite{Vaitaitis:1999wq}, E949~\cite{Artamonov:2014urb}, MicroBooNE~\cite{Abratenko:2019kez}, T2K~\cite{Abe:2019kgx}, ATLAS~\cite{Aad:2019kiz}, Belle~\cite{Liventsev:2013zz}, and Big Bang Nucleosynthesis (BBN)~\cite{Boyarsky:2020dzc}\footnote{We cut off the BBN constraints above $\left|U\right|=10^{-5}$ to match the information presented in Ref.~\cite{Boyarsky:2020dzc}, but naturally expect the limits to extend above this range.} (see e.g., Ref.~\cite{Beacham:2019nyx} for a thorough discussion of these limits).  For comparison, we also display the projected sensitivities to HNLs from several proposed experiments, including 
 NA62++\cite{Drewes:2018gkc}, FASER~\cite{Kling:2018wct}, CODEX-b~\cite{Aielli:2019ivi}, MATHUSLA~\cite{Curtin:2018mvb} and SHiP~\cite{Alekhin:2015byh}. 
For additional proposals to probe GeV-scale HNLs see e.g., Refs.~\cite{Ballett:2016opr,Ballett:2019bgd,Beacham:2019nyx,Berryman:2019dme,Coloma:2019htx,Coloma:2020lgy,Hirsch:2020klk}.  

We conclude that DarkQuest Phase I can probe a significant region of currently unexplored parameter space for $\tau$-aligned HNLs. For the Phase II scenario, DarkQuest will be able to extend the sensitivity by more than one order of magnitude in the squared mixing angle compared to Phase I, while also covering new regions of parameter space in the $\mu$-aligned scenario which are presently unconstrained. 

\section{Dark Scalars}\label{Scalar}

We now consider dark scalars interacting through the Higgs portal. 
A new singlet scalar can couple to the SM Higgs through two renormalizable portal couplings, 
\begin{equation}
-{\cal L} \supset (A \hat S +\lambda \hat S^2) \hat H^\dag \hat H  .
\label{eq:Higgs-portal}
\end{equation}
The dark scalar may acquire a small coupling to SM fermions and gauge bosons through its mass mixing with the Higgs, which will occur if the $A\neq 0$ in (\ref{eq:Higgs-portal}) or if the dark scalar obtains a non-zero vacuum expectation value. 
Then, in the physical basis, the phenomenology at DarkQuest is governed by the dark scalar mass, $m_S$, and the scalar-Higgs mixing angle, $\theta$: 
\begin{equation}
{\cal L} \supset -\frac{1}{2}\,m_S^2 S^2 +  \theta \,   S \, \left( \frac{2m_W^2}{v} W_\mu^+ W^{\mu-} + \frac{m_Z^2}{v} Z_\mu Z^{\mu} - \sum_f \frac{m_f}{v} \bar f f    \right).
\label{eq:L-S}
\end{equation}
Given the experimental constraints on the mixing angle for dark scalars at the GeV-scale, we will always be working in the regime $\theta \ll 1$. 
We will not study the phenomenological consequences of additional couplings between the scalar and the Higgs, such as the cubic interaction $hSS$. While such a coupling can lead to additional scalar production processes such as $B\to KSS$, these are typically not as important at DarkQuest as processes involving singly produced scalars.  Such coupling also leads to Higgs exotic decays of the type $h\to SS$ \cite{Curtin:2013fra} that can be searched for at the LHC. We do not include the corresponding bounds in our summary plot in Fig. \ref{Fig:Scalar-phase1-production-channel}, since these bounds depend on the $hSS$ coupling that is independent from the mixing angle $\theta$.  
We now discuss in more detail the production of scalars, their decays, the experimental acceptance, and the DarkQuest reach.  

\subsection{Scalar production at DarkQuest}

At DarkQuest scalars are produced through three main processes:  
meson decays, proton bremsstrahlung, and gluon-gluon fusion. 
The sensitivity of DarkQuest to scalars produced through $B$ meson decays was already studied in Ref.~\cite{Berlin:2018pwi}. 
In this work we will also examine the potential additional sensitivity from scalars produced through kaon decays, proton bremsstrahlung, and gluon-gluon fusion. 

\begin{figure}
\begin{center}
\includegraphics[width=.5\textwidth]{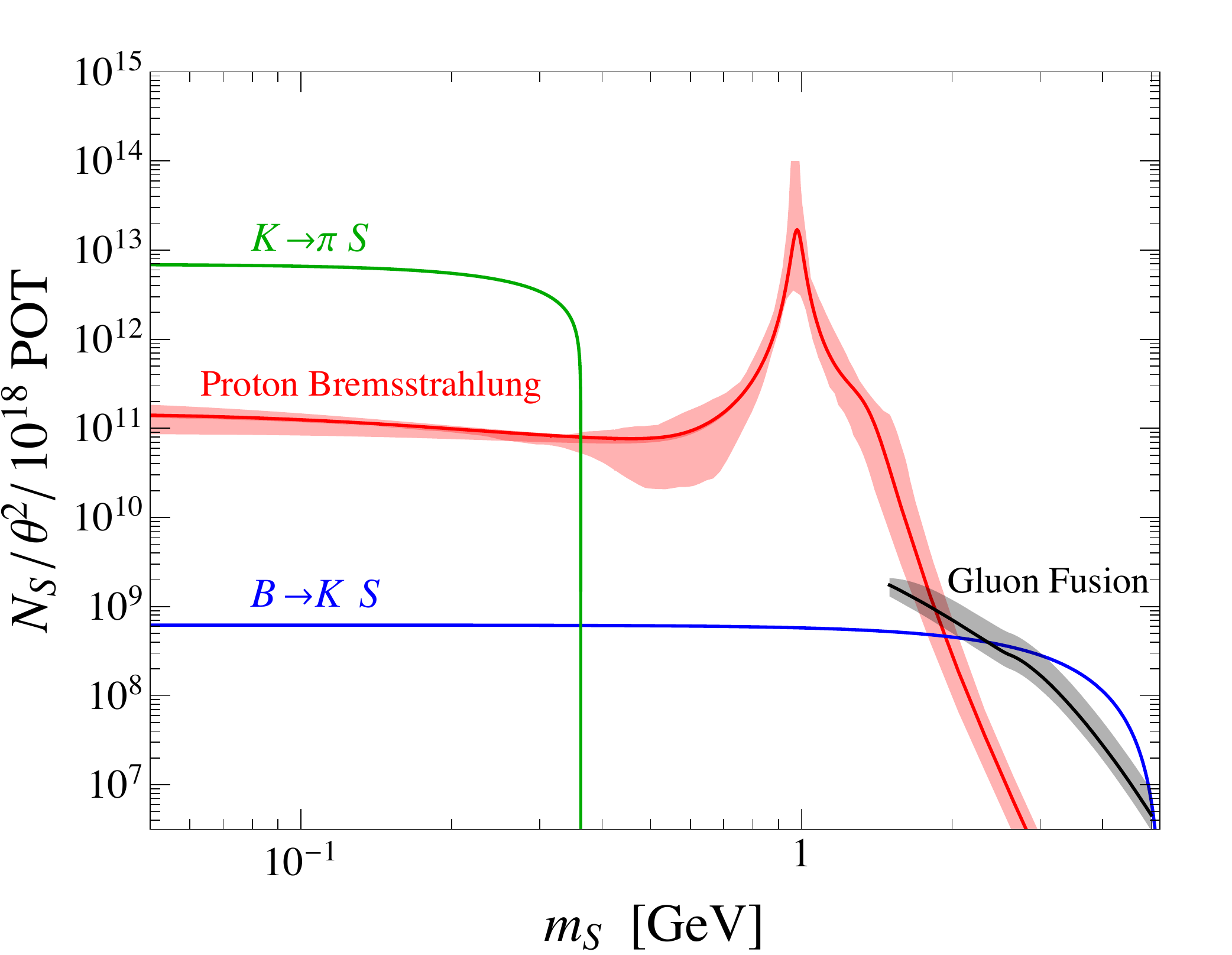}
\end{center}
\caption{Number of scalars produced at DarkQuest for $K \rightarrow \pi S$ (green), $B\rightarrow K S$ (blue), proton bremsstrahlung (red), and gluon fusion (black),  assuming $10^{18}$ protons on target and a mixing angle equal to 1. \label{Fig:NScalar}}
\end{figure}

Figure~\ref{Fig:NScalar} shows the number of dark scalars produced through these three production channels as a function of the scalar mass, assuming  $10^{18}$ protons on target.
 Low mass scalars are dominantly produced in kaon decays. Above the $m_K-m_\pi$ threshold and in the vicinity of $m_S \sim 1$ GeV, proton bremsstrahlung dominates, while heavier scalars can be produced through $B$-meson decays and  gluon fusion.

\subsubsection{Meson decays}

We first consider scalar production through meson decays. We refer the reader to Sec.~\ref{sec:particlesatDQ} and Table~\ref{table:SigmaMeson} for a summary of meson production at the DarkQuest.  
We first consider scalars produced through kaon decays, $K \rightarrow \pi S$, which is especially relevant for lighter scalars. 
The partial decay width for $K^\pm \rightarrow \pi^\pm S$ is~\cite{Willey:1982mc,Leutwyler:1989xj,Bezrukov:2009yw,Kamenik:2011vy,Winkler:2018qyg}
\begin{equation}
\Gamma(K^\pm \rightarrow \pi^\pm S) \simeq  
\frac{ \theta^2}{16 \pi m_K}  \bigg\vert \frac{3 G_F \sqrt{2} V^*_{td} V_{ts} m_t^2 m_s }{16 \pi^2 v}  \bigg\vert^2 \left(\frac{1}{2}\frac{m_K^2 - m_\pi^2}{m_s-m_d} f_K \right)^2 
 \lambda^{1/2}\left(1, \frac{m_S^2}{m_K^2}, \frac{m_\pi^2}{m_K^2}\right), \\
\end{equation} 
with $\Gamma(K^0_L \rightarrow \pi^0 S) \simeq \Gamma(K^\pm \rightarrow \pi^\pm S)$.%
\footnote{Although the branching fractions are different, the partial widths are very similar, and the total width cancels out of the estimate (\ref{eq:NKs}) as long as $\lambda_K \ll \langle \gamma_K^{-1} \rangle c \tau_i$.  In fact, $K^0_S$ has $\lambda_K \sim \langle \gamma_K^{-1} \rangle c \tau_{K_S}$ suggesting its total width could also cancel out of the expression (up to an $\mathcal O(1)$ factor).  However, $K_S^0$ is not included in our analysis since the partial width $\Gamma(K^0_S \rightarrow \pi^0 S) \ll  \Gamma(K^0_L \rightarrow \pi^0 S)$, so it can be neglected for that reason.} 
Using these partial widths and (\ref{eq:NKs}), the number of scalars produced from kaon decays in a thick target can be estimated as~\cite{Winkler:2018qyg} 
\begin{equation}
N_S  = N_p \, n_K \, \Gamma(K\rightarrow \pi S)  \, \lambda_K \langle\gamma_K^{-1} \rangle \,
\sim   10^{13} \times  \theta^2 \, \left( \frac{N_p}{10^{18}} \right)   ~~~~~~~({\rm kaon~decays}), ~~~
\label{eq:NS-K-decay} 
\end{equation}
where $n_K \sim 0.6$ is the number of $K^\pm$ and $K_L^0$ produced per proton on target.

Next, we consider scalars produced through $B$ meson decays, which proceeds through $b-s-S$ penguin transitions.
The inclusive branching ratio for $B \rightarrow X_s S$ can be written as~\cite{Evans:2017lvd,Gligorov:2017nwh,Grinstein:1988yu} (see also \cite{Batell:2009jf,Winkler:2018qyg} for exclusive $B$ decays)
\begin{equation} \label{eq: Bdecay}
\frac{ {\rm Br} (B\rightarrow X_{s} \, S) }{  {\rm Br} (B\rightarrow X_{c} \, e\, \nu ) } 
 \simeq  \theta^2 \, \frac{27 \sqrt{2} \,G_F \, m_t^4}{64 \pi^2 \, \Phi \, m_b^2} 
  \bigg\vert  \frac{V^*_{ts} V_{tb}}{V_{cs}}  \bigg\vert^2     \left(1-\frac{m_S^2}{m_b^2}\right)^{2},
\end{equation}
where $\Phi \approx 0.5$ is a phase space factor. Using the measured inclusive rate for  $B\rightarrow X_{c} \, e \, \nu$~\cite{Tanabashi:2018oca}, we obtain
${\rm Br} (B\rightarrow X_{s} \, S) \simeq 6.2 \times \theta^{2} (1-m_S^2/m_b^2)^{2}$. 
Since $B$-mesons decay promptly, we can estimate the number of scalars produced in their decays as
\begin{eqnarray}
N_S  =  N_p \, n_B \, {\rm Br}(B \rightarrow X_s \, S)   \sim  10^9 \times  \theta^2 \, \left( \frac{N_p}{10^{18}}   \right)    ~~~~~~~~~~(B~{\rm meson~decays}),
\label{eq:NS-B-decay}
\end{eqnarray}
where $n_B \sim  10^{-10}$ is the number of $B$ mesons produced per proton on target at DarkQuest.

\subsubsection{Proton bremsstrahlung}

Next, we turn to scalars produced through proton bremsstrahlung, $p+p \rightarrow S +X$. The cross section is obtained following the calculation in Ref.~\cite{Boiarska:2019jym}, which is based on the generalized Weizsacker-Williams method~\cite{Kim:1973he}; further details are provided in Appendix \ref{Appendix:brem}. Specifically, scalar events are generated by sampling the differential cross section $d \sigma_{\rm brem}/dz \, dp_T^2$, where 
$z \equiv p_S/p_p$ is the fraction of the proton beam momentum, $p_p$, carried by the emitted scalar, with $p_S$ the scalar momentum, and $p_T$ is the scalar transverse momentum. 
The validity of the Weizsacker-Williams approach relies on the kinematic conditions 
$p_p$, $p_S$, $p_p - p_S \gg m_p$, $|p_T|$. To satisfy these conditions for DarkQuest that uses 120 GeV protons, 
we follow Ref.~\cite{Berlin:2018pwi} and restrict the phase space to the range $z\in (0.1,0.9)$ and $p_T < 1$ GeV. We note that these conditions are slightly more restrictive than those used in Ref.~\cite{Boiarska:2019jym}, leading to an integrated cross section that is smaller by an order one factor. 

The total bremsstrahlung cross section is estimated to be
\begin{equation}
\sigma_{\rm brem} \sim  \, \sigma_{pp} \times \left( \frac{g_{SNN}^2 \theta^2 }{8 \pi^2}   |F_S(m_S^2)|^2   \right) ,
\end{equation}
where $\sigma_{pp} \approx 40$ mb is the total inelastic proton-proton cross section and the factor in parentheses gives the approximate integrated probability of scalar emission. The parameter
$g_{SNN}$ is the zero momentum scalar nucleon coupling (for $\theta = 1$) and  $F_S(p_S^2)$ is a time-like scalar-nucleon form factor, which is discussed in detail in Appendix \ref{Appendix:brem}.
Including order one factors arising from phase space integration, we estimate the total number of scalars produced in proton bremsstrahlung to be
\begin{equation}
N_S \sim 10^{11} \, \theta^2 \,  \left( \frac{N_p}{10^{18}}  \right)  ~~~~~~~~~~ ({\rm Proton ~ Bremsstrahlung}).
\label{eq:NS-Brem}
\end{equation}
Figure~\ref{Fig:NScalar} shows the total number of scalars produced at DarkQuest as a function of the scalar mass. The large resonant enhancement near $m_S \sim 1$ GeV is a consequence of mixing with the narrow $f_0(980)$ scalar resonance, while the bremsstrahlung cross section drops steeply for $m_S \gtrsim 1$ GeV due to the form factor suppression. 
It is likely that the zoo of heavy $f_0$ resonances would belay this high mass suppression, but we make no attempt to model that here. The uncertainty band is obtained by varying the lower integration limit for $z$ between 0.05 and 0.2 as well as the scalar resonance masses and widths in the form factor $F_S(p_S^2)$.

We note that the rates for scalar production from bremsstrahlung have a rather mild dependence on the proton beam energy, and thus the production rate at higher energy facilities such as the CERN SPS (400 GeV protons) is very similar to that at DarkQuest. 

\subsubsection{Gluon fusion}

The final process we consider is scalar production via gluon fusion. As in the case of the SM Higgs boson, this process proceeds at one loop through the heavy quark triangle diagrams. The full leading order cross section is discussed in Appendix \ref{Appendix:gluonfusion}. We restrict our analysis to scalar masses above ${\cal O}(1 \, {\rm  GeV})$ where the perturbative QCD computation is valid. 
In this mass range, the cross section is of order $\sigma_{ggS} \sim 30 \, {\rm pb}  \times \theta^2 \,  (m_S / 1 {\rm GeV} )^{-2}$, and the number of scalars produced is therefore
\begin{equation}
N_{S} \sim 10^9  \times \theta^2 \, \left(\frac{1\, \rm GeV }{m_S}\right)^2 \left(\frac{N_p}{10^{18}} \right)~~~~~~~~~~ ({\rm Gluon ~ Fusion}).
\label{eq:NS-ggF}
\end{equation}
As in the case of the SM Higgs boson, we expect higher order corrections to enhance the rate by an order one factor, although we are not aware of an existing calculation in the literature that can be applied to such light scalars. While it would be interesting to study this question further, we will simply apply a $K$-factor equal to 1.5 in our estimate of the rate, which is similar to that of the SM Higgs boson. 
For our simulation, we use the HEFT model in {\textsc{MadGraph5$\_$}\lowercase{a}\textsc{mc$@$nlo}}~\cite{Alwall:2014hca} to generate scalar events, which are then passed to {\textsc{PYTHIA 8}}~\cite{Sjostrand:2007gs} for showering.
While we find that gluon fusion is generally subdominant to other production mechanisms (see the black curve in Figure~\ref{Fig:NScalar}), it can give some additional sensitivity in the 1-2 GeV scalar mass range, particularly for the Phase II scenario. For comparison, we find that the scalar production via gluon fusion is only about a factor of 2 larger at the higher energy CERN SPS.

\subsection{Scalar decays}

\begin{figure}
\begin{center}
\includegraphics[width=.49\textwidth]{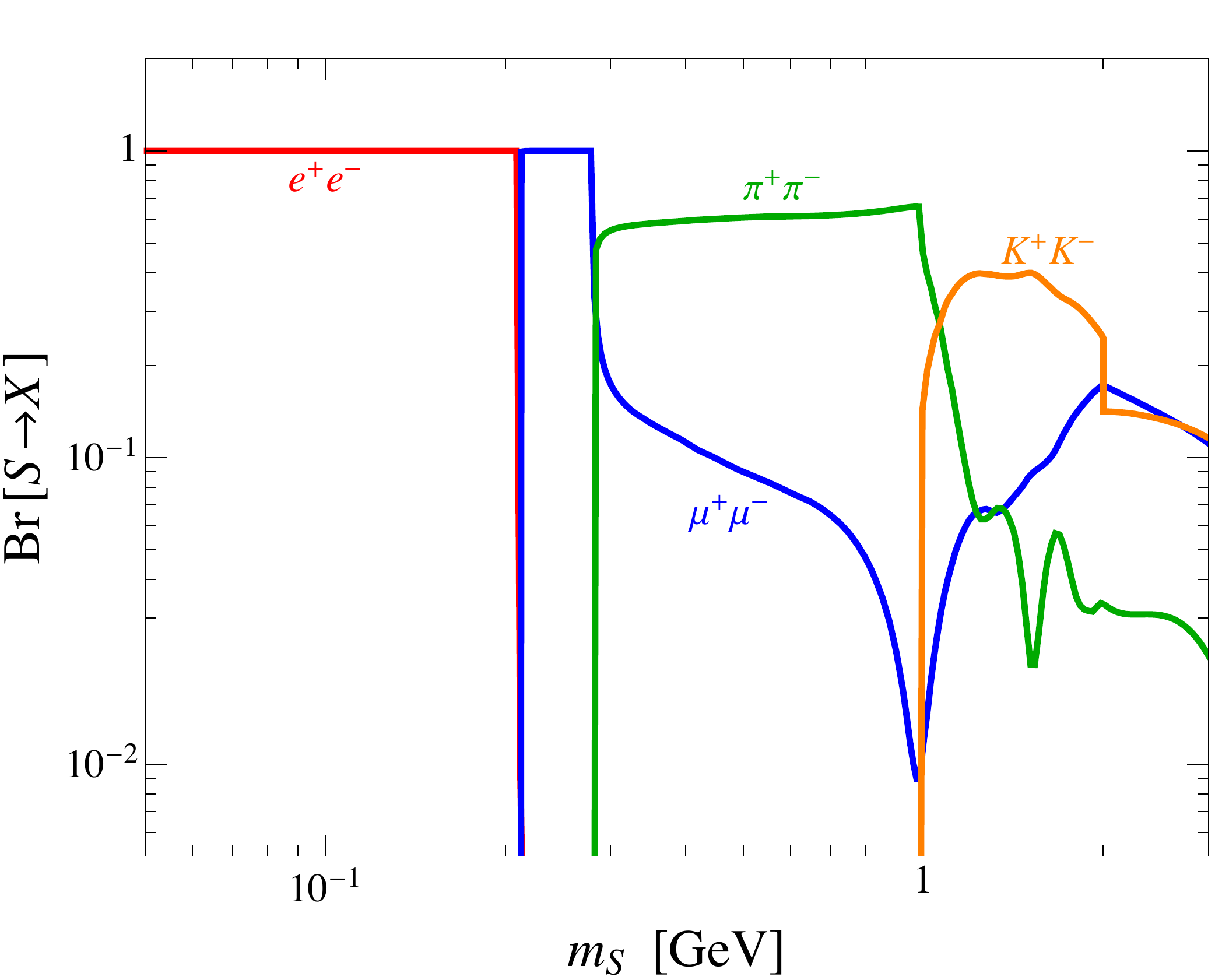}~~~~
\includegraphics[width=.49\textwidth]{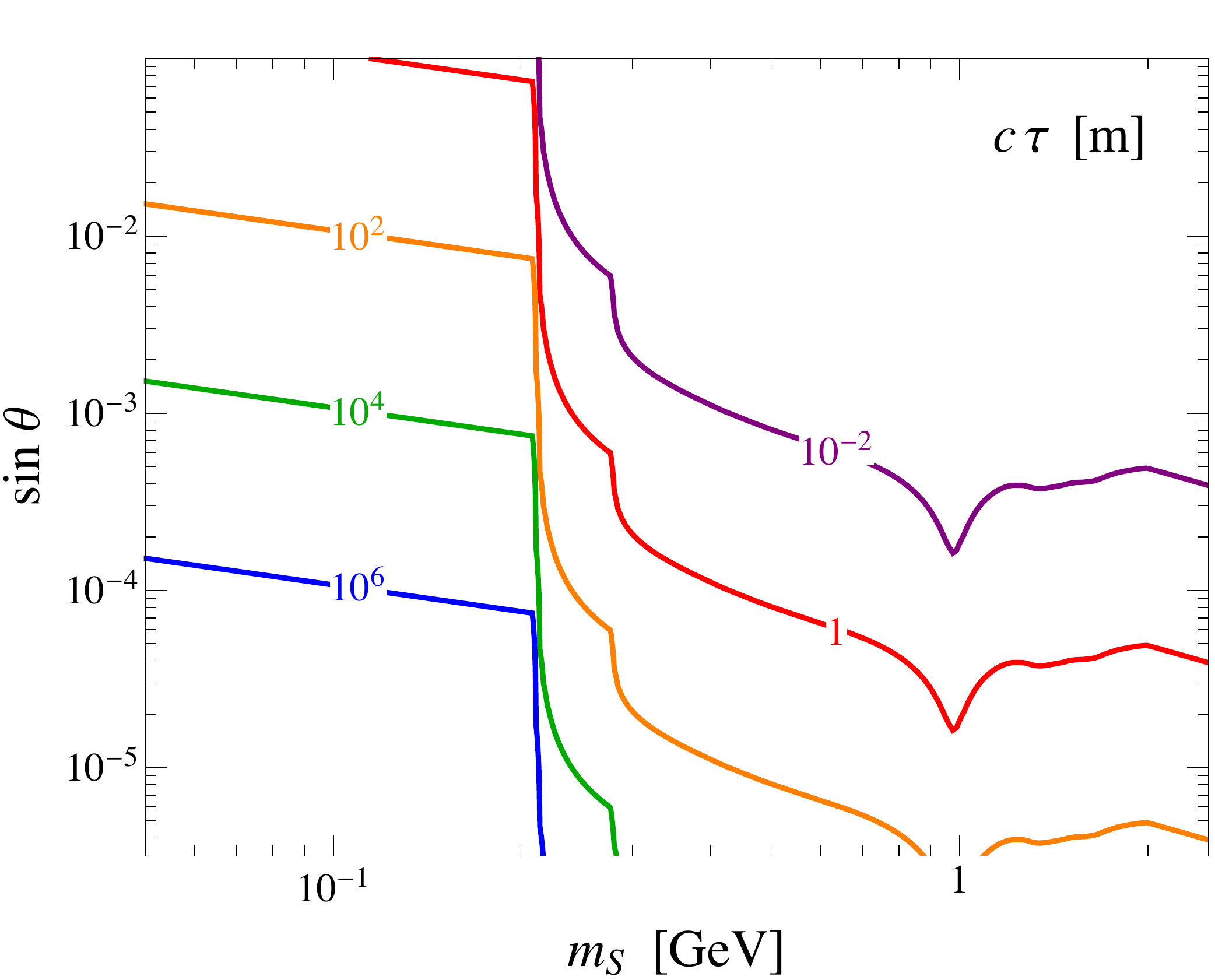}
\end{center}
\caption{ Left panel: Scalar branching ratios in the $e^+ e^-$ (red), $\mu^+ \mu^-$ (blue), $\pi^+ \pi^-$ (green), in the $K^+ K^-$ (orange) final state channels. Note that the branching ratios are independent of $\sin\theta$.
 Right panel:  Isocontours of the scalar decay length in units of meter in the $m_S$ - $\sin\theta$ plane. \label{Fig:BrS} }
\end{figure}

Through its mixing with the Higgs, the scalar will decay to SM final states. For example, the dark scalar can decay to charged leptons with a partial decay width, 
$\Gamma_{S \rightarrow \ell^+ \ell^-} \simeq \theta^2 m_\ell^2 m_S/ (8 \pi v^2)$.

Above the two pion threshold the scalar can also decay to hadronic final states. The theoretical description of such decays is complicated by strong interaction effects, leading to significant uncertainties in the predictions for masses of order 1 GeV. 
In our study we will use the results and prescriptions from the recent study in Ref.~\cite{Winkler:2018qyg}. In particular, for relatively low scalar masses  in the few hundred MeV range, the hadronic decays are well described using Chiral Perturbation Theory~\cite{Voloshin:1985tc,Donoghue:1990xh}.
At higher masses, $m_S \gtrsim 2$ GeV, the perturbative spectator model can be used to compute the decay rates to quarks and gluons~\cite{Gunion:1989we}. In the intermediate regime of $m_S \sim 1-2$ GeV an analysis based on dispersion relations can be employed to estimate the partial decay widths for scalar decays to pairs of pions and kaons~\cite{Raby:1988qf,Truong:1989my,Donoghue:1990xh,Monin:2018lee,Winkler:2018qyg}. Furthermore, Ref.~\cite{Winkler:2018qyg} includes an additional contribution to the scalar decay width to account for other hadronic channels above the 4$\pi$ threshold.
Despite the formidable calculations involved in estimating the decays in these regimes, these are uncontrolled approximations and should be viewed with healthy skepticism \cite{Bezrukov:2018yvd}.
 The scalar branching ratios in the $e^+ e^-$, $\mu^+ \mu^-$, $\pi^+ \pi^-$, and $K^+ K^-$ channels, as well as the scalar decay length, are shown in Figure~\ref{Fig:BrS}. 

As with our HNL projections presented in Sec.~\ref{RHNReach}, we will require 10 signal events in our dark scalar sensitivity estimates. 
The considerations leading to this assumption are similar to those outlined in Secs.~\ref{sec:acceptance} and \ref{RHNDecays}. In particular, for the signatures arising from scalar decays to leptons, $S\rightarrow \ell^+ \ell^-$, there can be backgrounds from  $K_L^0$ that pass through the FMAG and decay via $K_L^0 \rightarrow \pi^\pm \ell^\pm \nu$, though we expect that detector level pion-lepton discrimination can be used to bring these backgrounds at the level of $\mathcal O(10)$ ($< 1$) events for Phase I (Phase II). 
For the hadronic scalar signatures such as $S\rightarrow \pi^+ \pi^-, K^+ K^-$,  there are backgrounds from the decays $K_L^0 \rightarrow \pi^- \pi^+ \pi^0$ and $K_L^0\rightarrow \pi^+ \pi^-$. 
 The corresponding background rates, particularly for the two pion decay,  are further suppressed by the small branching ratios (BR$(K_L^0\to\pi^+\pi^-)\sim 2\times 10^{-3})$, and we expect that kinematic information will be helpful in distinguishing the signal, though this remains to be studied in detail.

\subsection{Detector acceptance}\label{Sec:ScalarAcceptance}

\begin{figure}
\begin{center}
\includegraphics[width=.5\textwidth]{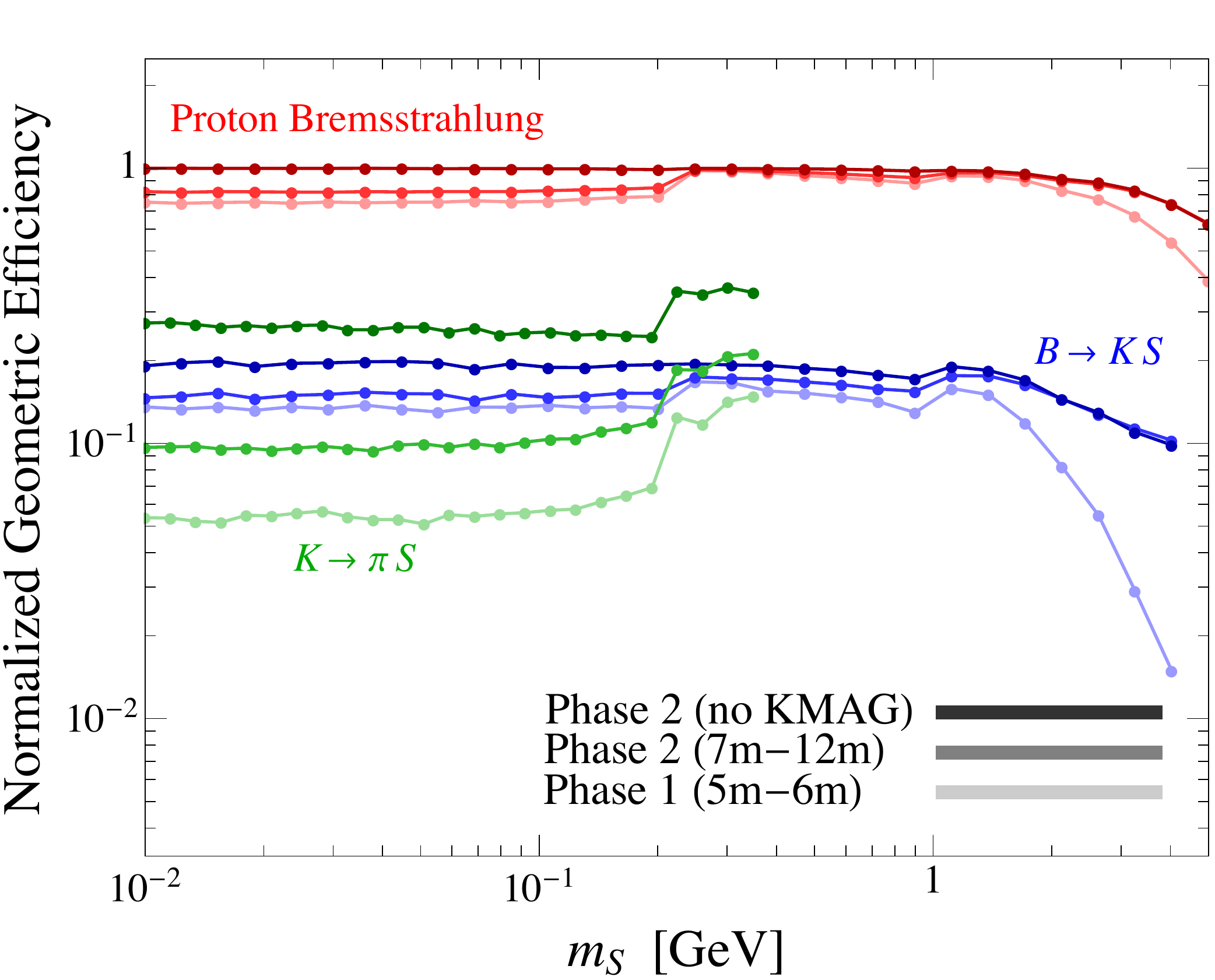}~~~~
\end{center}
\caption{Geometric acceptance as a function of scalar mass normalized to the number of scalars decaying within the fiducial decay region in the infinite lifetime limit. We show separately the efficiency for scalars produced via proton bremsstrahlung (red), $B$ decays (blue), and kaon decays (green), and for three run scenarios: Phase I, 5 m $-$ 6 m (light shading), Phase II, 7 m $-$ 12 m, (medium shading) and Phase II, 7 m $-$ 12 m, without the KMAG (dark shading).  
The acceptance combines the $e^+ e^-$, $\mu^+ \mu^-$, $\pi^+ \pi^-$, and $K^+ K^-$ final states weighted by their relative decay rates. \label{Fig:scalar-eff}}
\end{figure}

We follow the procedure discussed in Sec.~\ref{sec:acceptance} to account for the geometric acceptance of the experiment, with the total detector efficiency  computed according to Eq.~(\ref{eq:efficiency}).

In Figure~\ref{Fig:scalar-eff} we display the geometric acceptance as a function of scalar mass in the infinite lifetime limit, normalized to the number of scalars decaying within the fiducial decay region. 
This limit is of practical importance for much of the small $\theta$ parameter space. 
Several notable features can be observed in Figure~\ref{Fig:scalar-eff}. 
First, the overall efficiency is higher for dark scalars produced in proton bremsstrahlung compared to those from $B$ and kaon decays. This is due to the larger typical Lorentz boosts of scalars originating in the former process, which inherit an order one fraction of the beam energy. 
Second, an increase in the efficiency is typically observed as $m_S$ increases beyond the dimuon threshold. 
Due to phase space suppression, heavier particles produced through scalar decays will typically be more collinear with the parent scalar, which leads to a higher overall acceptance. 
Furthermore, in the decays to electrons, the emitted particles are highly relativistic in the scalar rest frame and the fraction emitted towards the negative $z$ direction can have a small lab frame longitudinal momentum. 
Such electrons can be swept out of the detector as they pass through the KMAG, explaining in the lower observed efficiency when the KMAG is present. 
Furthermore, we see that for heavy scalars produced via bremsstrahlung and $B$-meson decays, the efficiency tends to decrease as the the scalar mass increases beyond ${\cal O}(1\, {\rm GeV})$ since in this regime the daughter particle $p_T$ inherited from the scalar mother increases approximately in proportion to $m_S$ and is generally larger than that imparted by the KMAG. 
Another trend observed in all production channels is the increased efficiency in Phase II (medium shading) over that in Phase I (lighter shading), which stems from the fact that for the Phase II scenario the scalars decay closer to tracking station 3.

 Finally, we have displayed the efficiency for an alternate Phase II scenario in which the KMAG is removed and the charged daughters are not deflected. 
In this case, the daughter particles have a smaller characteristic transverse momentum, leading to a higher geometric acceptance as seen in Figure~\ref{Fig:scalar-eff}. 
However, it should also be emphasized that in this run scenario particle momenta measurement capability is likely to be significantly degraded. 
In fact, the magnetic field strength of the KMAG is tunable~\cite{KMAG-tunable} and could impart a smaller $p_T$ kick than the 0.4~GeV used in this work.  It would be interesting to study in detail its impact on the geometric acceptance and reconstruction capabilities.

\subsection{DarkQuest sensitivity to dark scalars}\label{Sec:reachScalars}

\begin{figure}[t!]
\begin{center}
\includegraphics[width=.5\textwidth]{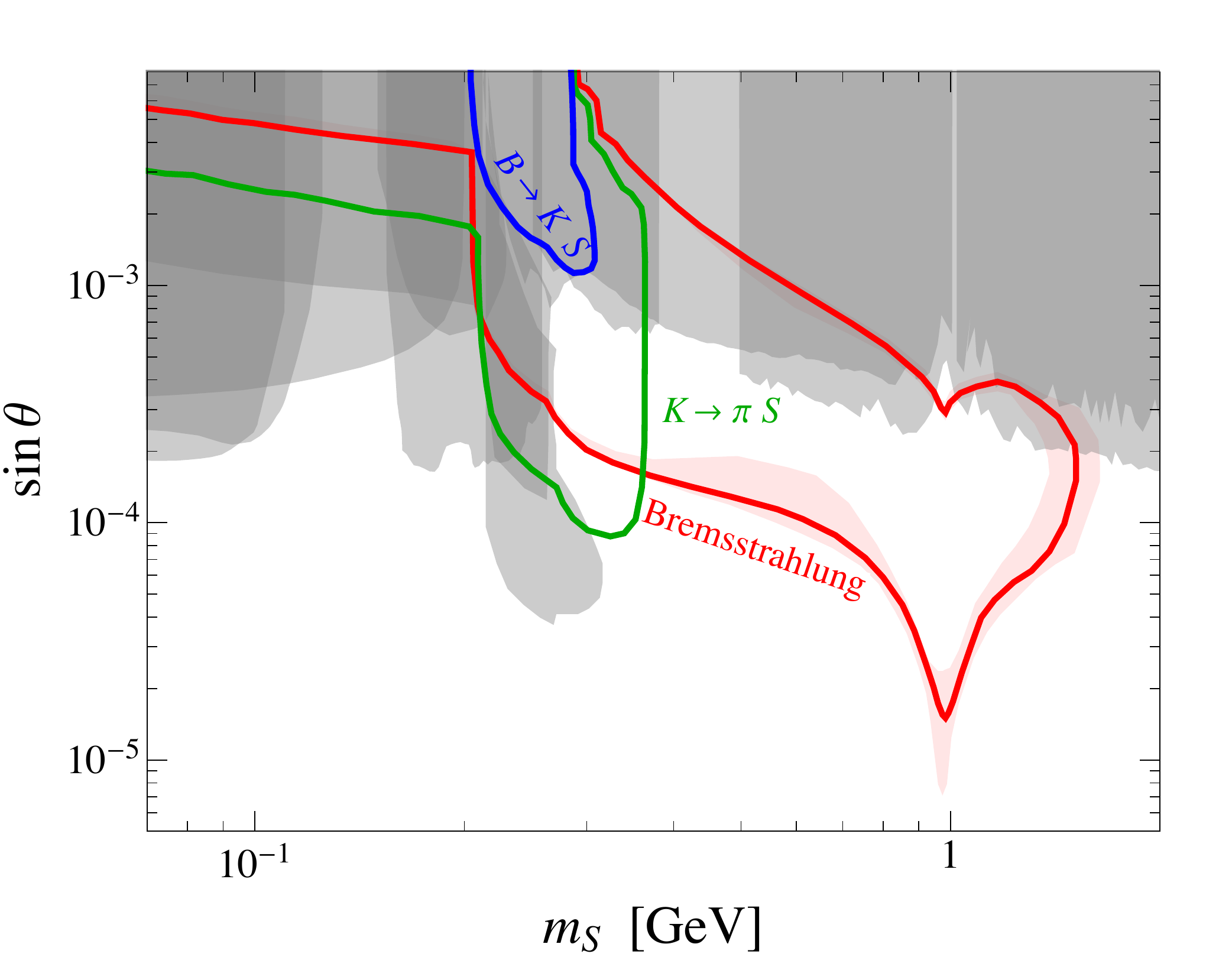}
\end{center}
\caption{DarkQuest Phase I sensitivity to dark scalars corresponding to $N_p=10^{18}$  and 5 m - 6 m decay region. The contours correspond to 10 signal events as obtained by adding the $e^+ e^-$, $\mu^+ \mu^-$, $\pi^+ \pi^-$, $K^+ K^-$ channels, for dark scalars produced via $K \rightarrow \pi S$ (green), $B\rightarrow K S$ (blue), and proton bremsstrahlung (red). The gray shaded regions correspond to existing limits from past experiments; see text for further details. \label{Fig:Scalar-phase1-production-channel}}
\end{figure}
Given the scalar production rates, decay branching ratios, lifetime, and experimental efficiency, 
we can now estimate the total number of signal events in the SM final state $i$ according to the formula 
\begin{equation}
N_{\rm signal} = N_S \times {\rm Br}_i \times {\rm eff}_i \, , 
\label{eq:NS}
\end{equation}
where $N_S$ is the number of scalars produced in a given production channel (see Eqs.~(\ref{eq:NS-K-decay}, \ref{eq:NS-B-decay}, \ref{eq:NS-Brem}, \ref{eq:NS-ggF}) for the number of scalars produced via $K$ decay, $B$ decay, bremsstrahlung, and gluon fusion, respectively).  
In Figure~\ref{Fig:Scalar-phase1-production-channel} we show the projected per-production-channel 
sensitivity of DarkQuest Phase I for scalars decaying inclusively to pairs of charged particles, specifically $e^+ e^-$, $\mu^+ \mu^-$, $\pi^+ \pi^-$, and $K^+ K^-$. Each contour indicates the scalar mass - mixing angle parameters predicting 10 signal events according to (\ref{eq:NS}).  
We show three contours corresponding to distinct scalar production mechanisms, including kaon decays, $B$-meson decays, and proton bremsstrahlung. 
No sensitivity is obtained from the gluon fusion process alone in the Phase I run scenario. 
The gray shaded regions indicate parameter points that are excluded by past experiments, which will be discussed in more details below. 
We observe from Figure~\ref{Fig:Scalar-phase1-production-channel} that DarkQuest Phase I (5m - 6m, $N_p=10^{18}$) will be able to explore a significant new region of parameter space, 
in particular for scalars produced through kaon decays and proton bremsstrahlung.

\begin{figure}[h]
\begin{center}
\includegraphics[width=.5\textwidth]{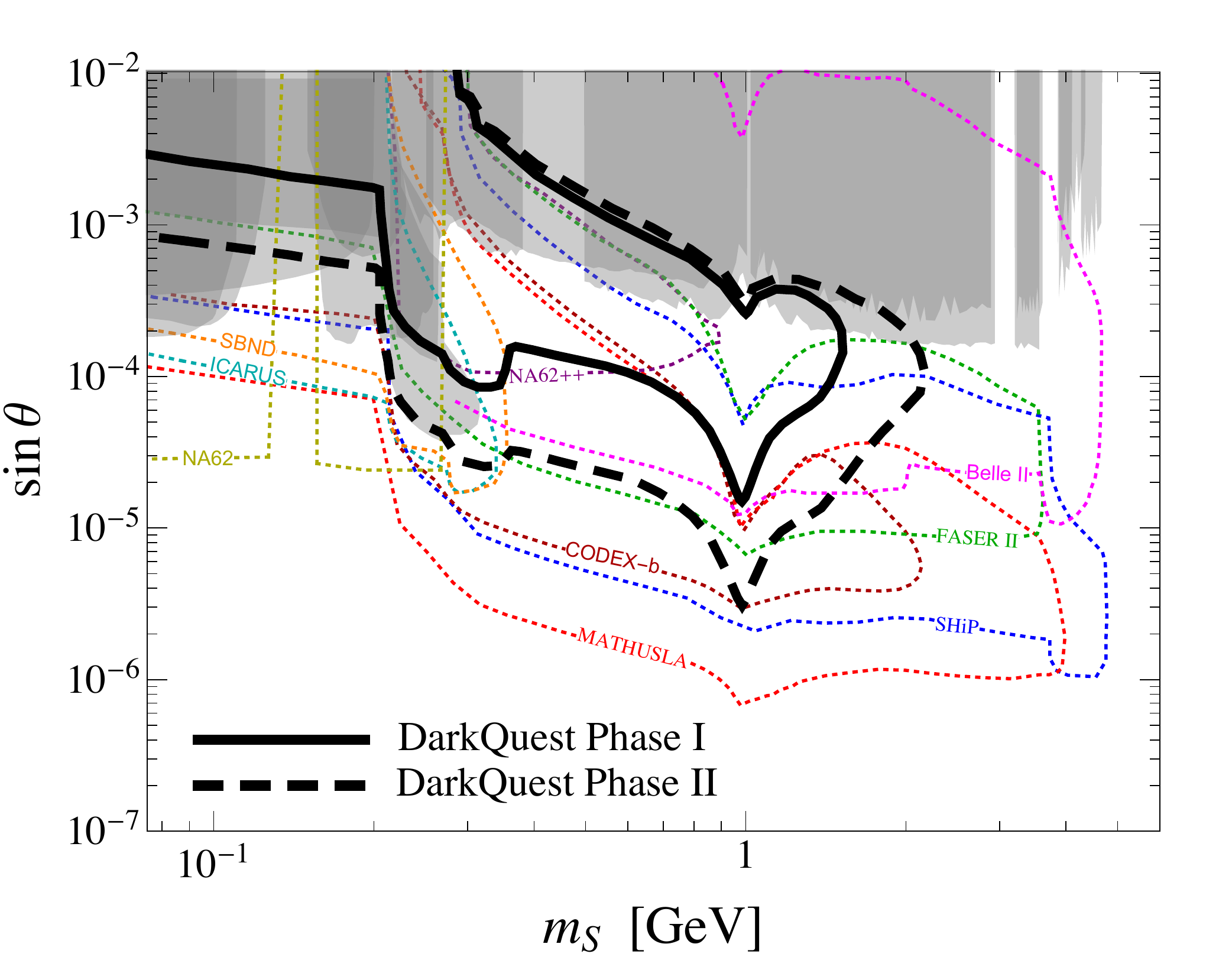}
\end{center}
\caption{
DarkQuest sensitivity to dark scalars. The contours correspond to 10 signal events as obtained by adding the $e^+ e^-$, $\mu^+ \mu^-$, $\pi^+ \pi^-$, $K^+ K^-$ channels, for combined dark scalar production via $K \rightarrow \pi S$, $B \rightarrow K S$,  proton bremsstrahlung and gluon fusion. We display both the DarkQuest Phase I sensitivity (solid, black) corresponding to $N_p=10^{18}$  and 5 m - 6 m decay region, as well as the DarkQuest Phase II sensitivity (dashed, black) corresponding to $N_p=10^{20}$  and 7 m - 12 m decay region. The gray shaded regions correspond to existing limits from past experiments. Also displayed are estimates from a variety of proposed experiments; see the text for further details and discussion.  \label{Fig:Scalar-P1-P2-plots}
}
\end{figure}

Next, in Figure~\ref{Fig:Scalar-P1-P2-plots} we show the full DarkQuest sensitivity to scalars decaying inclusively to pairs of charged particles, now combining all $S$ production channels, for both Phase I (solid, black) and Phase II (dashed, black) scenarios.  
In comparison to Ref.~\cite{Berlin:2018pwi}, which studied scalars produced only in $B$-decays, we find that the additional scalar production  from kaon decays and proton bremsstrahlung can significantly expand the parameter space that can be probed by DarkQuest.\,\footnote{We have compared our projections with Ref.~\cite{Berlin:2018pwi} for scalars produced via $B$ decays and find good agreement.}
In the figure, we also show the current experimental bounds on dark scalar parameter space, including those from CHARM~\cite{Bergsma:1985qz,Winkler:2018qyg}, LSND~\cite{Foroughi-Abari:2020gju}, E787/E949~\cite{Artamonov:2008qb,Artamonov:2009sz}, LHCb~\cite{Aaij:2016qsm,Aaij:2015tna}, and NA62~\cite{NA62limit}. 
In addition, we also display sensitivity projections from several ongoing or proposed future experiments, including NA62~\cite{Bondarenko:2019vrb,Beacham:2019nyx}, SBND and ICARUS~\cite{Batell:2019nwo}, Belle II~\cite{Kachanovich:2020yhi} (see also Ref.~\cite{Filimonova:2019tuy}), FASER~\cite{Feng:2017vli}, CODEX-b~\cite{Gligorov:2017nwh}, MATHUSLA~\cite{Curtin:2018mvb} and SHiP~\cite{Alekhin:2015byh}. 
See also e.g., Refs.~\cite{Beacham:2019nyx,Berryman:2019dme,Moulson:2018mlx,Archer-Smith:2020hqq} for further proposals to probe Higgs portal scalars in this mass range.\footnote{We also note that a recent excess observed by the KOTO experiment can be explained in this scenario for scalar masses $m_S \sim 150$ MeV and mixing angles $\theta \sim {\rm few} \times 10^{-4}$~\cite{Egana-Ugrinovic:2019wzj}.} We observe that DarkQuest Phase I has the potential to cover a significant region of unexplored parameter space for scalar masses between about  200 MeV and 2 GeV.  Phase II will probe angles as small as $\theta \gtrsim  5 \times 10^{-6}$ and as large as $ \theta \lesssim 10^{-3}$. 

\section{Summary}\label{sec:summary}

We have investigated the sensitivity of the Fermilab DarkQuest experiment  to two simple and well-motivated dark sector scenarios, heavy neutral leptons and Higgs-mixed scalars. The proposed DarkQuest ECAL upgrade will allow for sensitive searches to a variety of displaced final states containing charged particles and photons, which arise in the models considered here from the decay of long lived HNLs or scalars. We have carefully estimated the production and decay rates of these dark sector particles as well as the detector acceptance to derive projections under two benchmark run scenarios. 
During the Phase I scenario based on $10^{18}$ protons on target and a 5m - 6m fiducial decay region, DarkQuest will be able to explore significant new parameter space for $\tau$-mixed HNLs and dark scalars in the mass range of a few hundred MeV - 2 GeV. It is conceivable that this could be achieved on the 5 year time scale, putting DarkQuest on a competitive footing with other proposed experiments. 
Looking down the road, a potential Phase II scenario with $10^{20}$ protons on target and a 7m-12m fiducial decay region would allow for improvements by more than one order of magnitude in terms of the interaction rates with SM particles (proportional to squared mixing angle). Our results build on past phenomenological studies~\cite{Gardner:2015wea,Berlin:2018tvf,Berlin:2018pwi,Choi:2019pos,Dobrich:2019dxc,Tsai:2019mtm,Darme:2020ral} and provide further motivation for the DarkQuest ECAL upgrade. 
This upgrade can be realized with a relatively modest investment and will leverage the existing experimental infrastructure already in place to build an exciting dark sector physics program at Fermilab.

\section*{Acknowledgements}

We thank Asher Berlin, Simon Knapen, Kun Liu, Nhan Tran, and Keping Xie for helpful discussions.
BB and MR are supported by the U.S. Department of Energy under grant No. DE- SC0007914.  JAE is supported by U.S. Department of Energy (DOE) grant DE-SC0011784. 
The research of SG is supported in part by the NSF CAREER grant PHY-1915852.


\section*{Appendix}
 \appendix
 \section{Direct Scalar Production}
\subsection{Proton Bremsstrahlung} \label{Appendix:brem}
To estimate the production rate of dark scalars  via proton bremsstrahlung, $p+p \rightarrow S +X$, we follow the calculation presented in Ref~\cite{Boiarska:2019jym} (see also Ref.~\cite{Foroughi-Abari:2020gju})
which employs the generalized Weizsacker-Williams method~\cite{Kim:1973he} to factorize the reaction to the two subprocesses: (i) emission of the scalar from the proton and (ii) proton-proton scattering. We denote the incoming proton momentum as $p_p$, the fraction of the proton beam momentum carried by the emitted scalar as $z= p_S/p_p$ with $p_S$ the scalar momentum, and the scalar transverse momentum as $p_T$. Provided the kinematic conditions, 
$p_p$, $p_S$, $p_p - p_S \gg m_p$, $|p_T|$
are satisfied, the differential production cross section can be factorized as 
\begin{equation}
\frac{d \sigma_{\rm brem}}{dz \, dp_T^2} \approx \sigma_{p p} (s') \, P_{p\rightarrow p S}(z, p_T^2).
\label{eq:brem-diff-XS}
\end{equation}
Here $\sigma_{p p}$ is the total $pp$ cross section, for which we use a fit to experimental data~\cite{Tanabashi:2018oca}, and $s' = 2\, m_p \, p_p \, (1-z) + 2 \,m_p^2$. The splitting probability for the scalar emission, 
$P_{p\rightarrow p S}(z, p_T^2)$, is computed using the old-fashioned perturbation theory approach~\cite{Altarelli:1977zs}:
\begin{equation}
P_{p\rightarrow p S}(z, p_T^2) \approx |F_{S} (m_S^2)|^2 \,\frac{g_{SNN}^2 \, \theta^2}{8 \pi^2} \,  \frac{z\, [m_p^2 \,(2-z)^2 + p_T^2]}{[m_p^2 \, z^2 + m_S^2 \,(1-z)+ p_T^2]^2},
\label{eq:brem-splitting}
\end{equation}
where $g_{SNN}$ is the scalar-nucleon coupling at zero momentum transfer, 
\begin{equation}
g_{SNN} = \frac{2}{9}\frac{m_N}{v} 
\left( 1+  \frac{7}{2} \sum_{q = u,d,s} \frac{m_q}{m_N}  \langle   N |    \bar q q  | N  \rangle \right) \approx 1.2 \times 10^{-3}.
\end{equation}
Furthermore, $F_{S}(p_S^2)$ in (\ref{eq:brem-splitting}) denotes the time-like form factor associated with the scalar-proton interaction. We will discuss our choice for this form factor below. To obtain the total cross section, (\ref{eq:brem-diff-XS}) is integrated over a restricted range of $z$, $p_T^2$ such that the kinematic conditions described above are satisfied. For our simulation of scalar production through proton bremsstrahlung, we generate scalar events with $z, p_T^2$ appropriately weighted according to the distribution in (\ref{eq:brem-diff-XS}).

\begin{figure}
\centering
\includegraphics[width=0.5\textwidth]{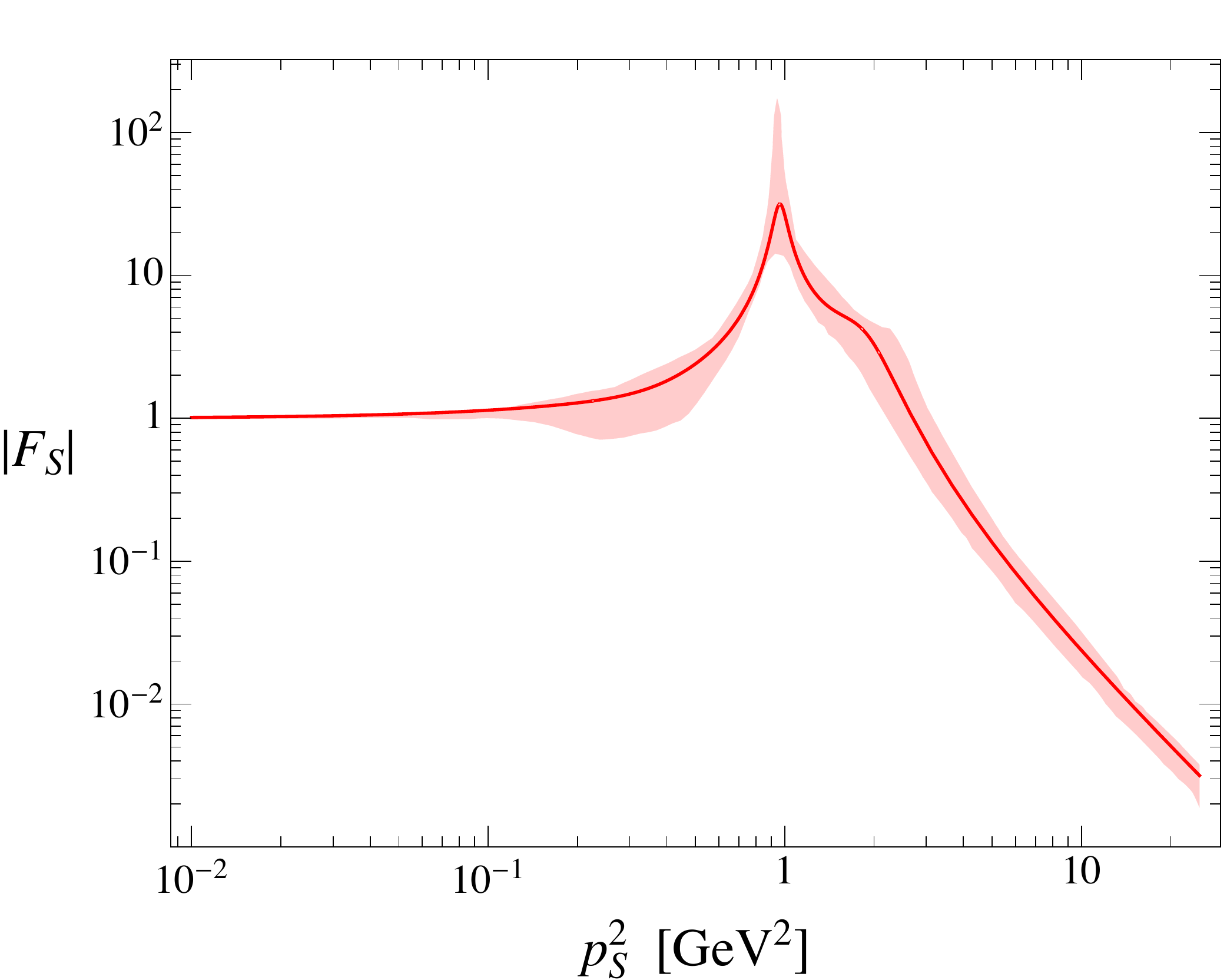}
\caption{Magnitude of time-like scalar form factor,  $|F_S(p_S^2)|$ from (\ref{eq:FS}). The central value (solid) is obtained for the mean values of the scalar resonance masses and widths. Varying the masses and widths within their quoted uncertainty range leads to the uncertainty band. 
\label{fig:form-factor}}
\end{figure}

We are not aware of any studies of the time-like scalar-nucleon form factor $F_S(p_S^2)$  in the literature. In analogy with vector meson dominance model of the time-like electromagnetic form factor discussed in Ref.~\cite{Faessler:2009tn} (commonly used for dark photon production via proton bremsstrahlung~\cite{Blumlein:2013cua,deNiverville:2016rqh}), we will assume that $F_S(p_S^2)$ incorporates mixing with $J^{PC} = 0^{++}$ scalar resonances through a sum of Breit-Wigner components,
\begin{equation}
F_S(p_S^2) = \sum_\phi \frac{f_\phi  \, m_\phi^2}{m_\phi^2 - p_S^2 - i \, m_\phi \, \Gamma_\phi},
\label{eq:FS}
\end{equation}
where we include the three low-lying scalar resonances,  $\phi \in \{f_0(500), f_0(980), f_0(1370)\}$. 
The decay constants $f_\phi$ for each resonance are obtained by imposing the conditions 
$F_S(0) =1$ and 
$F_S(p_S^2) \sim 1/{p_S^4}$  as  $p_S^2\rightarrow \infty$~\cite{Vainshtein:1977db}. A central value is defined by taking the mean values of the masses, $m_\phi = \{475,980,1350\}$ MeV, and widths, $\Gamma_\phi = \{550,55,350\}$ MeV, leading to the decay constants $f_\phi = \{280,1800,-990\}$ MeV. 
To provide a naive estimate of the uncertainty, we vary the masses and widths of the resonances within their quoted uncertainty ranges~\cite{Tanabashi:2018oca}, and take the envelope of the maximum and minimum values of $|F_S(p_S^2)|$. The magnitude of the form factor is plotted in Figure~\ref{fig:form-factor}. 

\subsection{Gluon Fusion} \label{Appendix:gluonfusion}

For scalars heavier than $\mathcal O(1~$ GeV$)$, one can consider perturbative QCD production processes. In analogy with the SM Higgs boson, the dominant production channel is gluon fusion, $gg \rightarrow S$. The production cross section can be written as 
\begin{equation}
\sigma_{ggS} = \theta^2 \frac{\alpha^2_S(\mu_R^2)}{1024 \, \pi \, v^2 }\bigg\vert \sum_q  A_{1/2}(\tau_q) \bigg\vert^2 {\cal L}_{gg}\left( \frac{m_S^2}{s}, \mu_F^2  \right),
\end{equation}
where $\tau_q  = m_S^2/4 m_q^2$,  $A_{1/2}$
 is a loop function (see e.g.,~\cite{Djouadi:2005gi}),
\begin{equation}
 A_{1/2}(\tau)=2 \left[\tau+(\tau -1 )f(\tau)\right] \tau^{-2},
\end{equation}
with $f(\tau)$ defined as 
\begin{equation}
f(\tau) = 
\begin{cases} 
\displaystyle{\arcsin^2 \sqrt{\tau} } ~~~~~~~~~~~~~~~~~~~~~~~~~~~~~~~~ \tau \leq 1\\
\displaystyle{-\frac{1}{4} \left[ \log \frac{1+\sqrt{1-\tau^{-1}}}{1-\sqrt{1-\tau^{-1} }} - i \pi \right]^2} ~~~~~~~ \tau >1.
\end{cases}
\end{equation}
Furthermore,  ${\cal L}_{gg}$ is the gluon parton luminosity function
\begin{equation}
{\cal L}_{gg}\left( \tau, \mu_F^2  \right) = \tau \int_\tau^1 \frac{dx}{x} f_g(x ,\mu_F^2) f_g(\tau/x,\mu_F^2),
\end{equation}
with $f_g(x)$ is the gluon PDF, and $\mu_R$ ($\mu_F$) is the renormalization (factorization) scale. 
To estimate the scale uncertainty in the cross section we fix $\mu_F=\mu_R=\mu$ and vary the scale between $\mu \,\in \, [\frac{2}{3}m_S,\frac{4}{3}m_S]$. 
Our projections in the gluon fusion channel are made with the  {\textsc{CT18NLO}}  PDF set \cite{Hou:2019efy} and use the  {\textsc{ManeParse}} package~\cite{Clark:2016jgm} for reading the PDF sets. 
We have also checked that our results do not change substantially under different choices of PDF sets. 
Since perturbative QCD breaks down at scales $Q \lesssim 1\text{ GeV}$, we only consider scalar production through gluon fusion for masses $m_{S}\gtrsim 1.5\text{ GeV}$.

\bibliographystyle{utphys}
\bibliography{SeaQuestBib}

\end{document}